\documentclass[twocolumn,usenatbib]{mn2e}

\usepackage{amsfonts}
\usepackage{amsmath}
\usepackage{amssymb}
\usepackage{aas_macros}
\usepackage{graphicx}
\usepackage{color}
\usepackage{float}

\def\lsim{~\rlap{$<$}{\lower 1.0ex\hbox{$\sim$}}}
\def\bsim{~\rlap{$>$}{\lower 1.0ex\hbox{$\sim$}}}

\def\hmpc{\ {\rm {\it h}^{-1}Mpc}}

\def\hmsun{\ {\rm M_\odot/{\it h}}}

\def\hhhmpc{\ {\rm {\it h}^{3}Mpc^{-3}}}

\def\cmm{\ {\rm cm^{-2}}}
\def\junit{\ {\rm ergs\,s^{-1}\,cm^{-2}\,sr^{-1}}}
\def\lunit{\ {\rm ergs\,s^{-1}}}

\def\dd{{\rm d}}
\def\ln{{\rm ln}}

\def\mathbi#1{\textbf{\em #1}}

\def\vx{\mathbi{x}}

\def\apjl{Astrophys.\ J.\ Lett.}

\def\aap{Astron.\ Astrophys.}

\def\apjs{Astrophys.\ J. Supp.}

\definecolor{RedWine}{rgb}{0.743,0,0}
\definecolor{RoyalBlue}{rgb}{0.25,.41,.88}
\definecolor{ForestGreen}{rgb}{.13,.54,.13}
\definecolor{DeepPurple}{rgb}{.72,.18,1}

\begin{document}

\title[UVB fluctuations with clustered sources]
{Ultraviolet background fluctuations with clustered sources}

\author[Desjacques, Moradinezhad Dizgah \& Biagetti]
{Vincent Desjacques\thanks{E-mail: Vincent.Desjacques@unige.ch}, 
  Azadeh Moradinezhad Dizgah and 
  Matteo Biagetti \\ 
  D\'epartement de Physique Th\'eorique and
  Center for Astroparticle Physics (CAP), Universit\'e de Gen\`eve, \\ 
  24 quai Ernest Ansermet, CH-1211 Gen\`eve, Switzerland}

\date{}
\label{firstpage}
\pagerange{\pageref{firstpage}--\pageref{lastpage}}

\maketitle

\begin{abstract}

We develop a count-in-cells approach to the distribution of ultraviolet background
fluctuations that includes source clustering. We demonstrate that an exact expression 
can be obtained if the clustering of ionising sources follows the hierarchical ansatz. 
In this case, the intensity distribution depends solely on their 2-point correlation 
function. We show that the void scaling function of high redshift mock quasars 
is consistent with the Negative Binomial form, before applying our formalism to the 
description of {\small HeII}-ionising fluctuations at the end of helium reionization. 
The model inputs are the observed quasar luminosity function and 2-point correlation 
at $z\sim 3$.
We find that, for an (comoving) attenuation length $\lesssim 55$ Mpc, quasar clustering 
contributes less than 30\% of the variance of intensity fluctuations so long as the 
quasar correlation length does not exceed $\sim 15$ Mpc. 
We investigate also the dependence of the intensity distribution on the large-scale 
environment. Differences in the mean {\small HeII}-ionising intensity between low- and 
high-density regions could be a factor of few if the sources are highly clustered. 
An accurate description of quasar demographics and their correlation with strong 
absorption systems is required to make more precise predictions.

\end{abstract}


\section{Introduction}
\label{sec:intro}

Modelling helium reionization is challenging because of the wide dynamical range that must be 
achieved to account simultaneously for the scarcity and clustering of the sources (quasars) 
and the physical properties of the low density intergalactic medium (IGM). 
Therefore, a number of hybrid methods combining analytic approaches with numerical simulations 
have been developed to address this problem
\citep[e.g.][]{sokasian/abel/hernquist:2002,gleser/nusser/etal:2005,bolton/haehnelt/etal:2006,
paschos/norman/etal:2007,furlanetto/oh:2008,fauchergiguere/lidz/etal:2009,meiksin/tittley:2012}.
Nevertheless, several issues, including the contribution of quasar clustering to the variance 
of the helium-ionising fluctuations towards the end of {\small HeII} reionization ($z\sim 3$), 
are still being debated. 
Whereas variations in the {\small HI}-ionising background are expected to be small owing to the 
large (comoving) attenuation length (or mean free path) of hydrogen-ionising photons, 
$r_0\sim 200$ Mpc \citep{prochaska/madau/etal:2014},
recent studies indicate that $r_0\sim 30 - 50$ Mpc only for helium-ionising
photons around $z\sim 3$ \citep{bolton/haehnelt/etal:2006,furlanetto/oh:2008,khaire/srianand:2013,
davies/furlanetto:2014}. 
This is not much larger than the observed clustering length $r_\xi\gtrsim 15 - 30$ Mpc of bright 
quasars in the same redshift range \citep[e.g.][]{shen/strauss/etal:2007,francke/gawiser/etal:2008}.
Clearly, $r_\xi/r_0\gtrsim 1$ is a necessary condition for source clustering to be important
However, the abundance of sources furnishes another characteristic length: the average source
separation  $l=\bar{n}^{-1/3}$. Hence, the condition $r_0/l\gg 1$ or, equivalently, a large number 
of sources per attenuation volume so that Poisson fluctuations are small relative to clustering 
effects, must also be satisfied. 
While bright quasars are very rare and, therefore, certainly do not meet this 
criterion, faint quasars are much more abundant, though possibly not as strongly clustered as their
bright companions. These considerations show that the importance of source clustering at the end 
of {\small HeII} reionization may strongly depends on the assumed quasar properties.

Recently, \cite{dixon/furlanetto/mesinger:2014} have addressed the impact of quasar clustering 
using a semi-numeric method, in which dark matter haloes are identified in realisations of the 
linear density field using the excursion set approach. They have found a relatively weak effect. 
However, given the current uncertainties on the demographics of high redshift quasars, it would 
be desirable to revisit this issue and explore a wider spectrum of quasar clustering amplitudes.

In this paper, we will investigate this issue analytically using an approach based on the 
count-in-cells formalism 
\citep[see e.g.][]{fall/geller/etal:1976,white:1979,peebles:1980,fry:1986a,balian/schaeffer:1989,
szapudi/colombi:1996}.
Our model generalises to clustered sources the early work of 
\cite{zuo:1992,fardall/shull:1993,meiksin/white:2003}, who 
considered the probability distribution of ionising intensity induced by randomly distributed 
sources. 
The assumption of hierarchical ansatz is a crucial ingredient of our method. It is efficient only 
if the source distribution follows the hierarchical scaling. 
We will show that this is the case of mock quasars at high redshift. 
This will enable us to explore very different quasar clustering configurations, at the expenses 
of a detailed modelling of the small-scale IGM physics. 

This paper is organised as follows. In Sec.\ref{sec:theory}, we introduce our count-in-cell 
approach, discuss the validity of the hierarchical ansatz for high redshift quasars and 
demonstrate that the intensity distribution $P(J)$ can be worked out exactly 
(within the simplifications of such analytic approaches) if the sources follows the hierarchical 
scaling. In Sec.\ref{sec:assymp}, we derive explicit 
scaling solutions for the low- and high-intensity tails and briefly discuss the numerical 
implementation of our result. In Sec.\ref{sec:results}, we apply our method to the distribution 
of {\small HeII}-ionising intensity at the completion of helium reionization. 
We discuss our results in Sec.\ref{sec:discussion} and conclude in Sec.\ref{sec:conclusion}.
We shall hereafter use $h=0.7$ in all unit conversions.

\section{Theoretical considerations}
\label{sec:theory}

The distribution of ionising intensities $P(J)$ has been worked out by 
\cite{zuo:1992,fardall/shull:1993,meiksin/white:2003} for Poisson distributed sources. 
Here, we extend their calculation to clustered sources. We begin with the introduction of 
position-dependent weights into the count-in-cells formalism before demonstrating that, if 
the sources follow the hierarchical scaling, then $P(J)$ can be recast into a simple 
expression.

\subsection{Cell counts with position-dependent weight}

Following \cite{white:1979}, we define the probability to have a cell of volume $V$ 
empty of particles except at positions $\vx_1$, ... , $\vx_N$ as 
\begin{equation}
P\bigl\{X_1\dots X_N\Phi_0(V)\bigr\} 
= P\bigl\{X_1\dots X_N\bigl\lvert\Phi_0(V)\bigr\} e^{{\cal W}_0(V)}\;.
\end{equation}
The probability $P_0 \equiv P(\Phi_0(V))$ to have an empty cell is the exponential of
the conditional void correlation \citep{fall/geller/etal:1976,white:1979,fry:1985}
\begin{align}
\label{eq:calW0}
{\cal W}_0(V) &= \sum_{k=1}^\infty \frac{(-\bar{n})^k}{k!}
\int_V\!\!\dd^3\!\vx_1\dots\int_V\!\!\dd^3\!\vx_k\,\xi_k(\vx_1,\dots,\vx_k) \\
&=\sum_{k=1}^\infty\frac{(-\bar{N})^k}{k!}\bar{\xi}_k(V) \nonumber \;.
\end{align}
Here, $\bar{n}$ is the average number density of objects, $\bar{N}=\bar{n}V$, 
$\xi_k(\vx_1,\dots,\vx_k)$ is the $k$-point irreducible correlation function and 
\begin{equation}
\bar{\xi}_k(V) \equiv \frac{1}{V^k}
\int_V\!\!\dd^3\!\vx_1\dots\int_V\!\!\dd^3\!\vx_k\,\xi_k(\vx_1,\dots,\vx_k)
\end{equation}
is its volume-average. Eq.(\ref{eq:calW0}) assumes that the volume can be split into 
many small sub-volumes, such that each individual cell is either empty or contains 
exactly one object. It would not hold if several objects could have the same location.
Note also that $\xi_1(\vx)\equiv 1$ for a homogeneous process. We will relax this 
assumption in Sec.\ref{sec:env}.

The void probability function $P_0$ is a generating function
for the count-in-cells probabilities.
Namely, the probability to have exactly $N$ objects in (randomly-located) cells of 
volume $V$ is 
\begin{equation}
\label{eq:PN}
P_N(V) = 
\frac{(-\bar{n})^N}{N!} \frac{d^N}{d\bar{n}^N}\exp[{\cal W}_0(V)] \;,
\end{equation}
where the derivatives are evaluated at constant $\bar{\xi}_k$ 
\citep{white:1979,sheth:1996}.
The positive definite, normalised probabilities $P_N(V)$ impose strong constraints on
the behaviour of ${\cal W}_0$ as a function of $V$ or, equivalently, $\bar{N}$
\citep[e.g.][]{fry:1985,balian/schaeffer:1989}. 
Clearly, we must have ${\cal W}_0(\bar{N})\leq 0$. Furthermore, the conditions 
$P_1>0$ and $P_2>0$ require
\begin{equation}
\frac{\partial{\cal W}_0}{\partial\bar{N}}>0 \qquad \text{and} \qquad
\frac{\partial^2{\cal W}_0}{\partial\bar{N}^2}
+\left(\frac{\partial{\cal W}_0}{\partial\bar{N}}\right)^2>0 \;.
\end{equation}
Assuming that the conditional void probability is locally of the form 
${\cal W}_0(\bar{N})=-\bar{N}^\beta$, this translates into the bound $0<\beta<1$. 
Finally, since we must recover the Poisson regime ${\cal W}_0(\bar{N})= -\bar{N}$ in 
the limit $\bar{N}\to 0$, this implies that ${\cal W}_0(\bar{N})$ is a convex, 
monotonically decreasing function of $\bar{N}$ that satisfies 
$-\bar{N}\leq {\cal W}_0(\bar{N})< 0$. In other words, $P_0$ is smallest for a Poisson
process.

For the purpose of modelling $P(J)$, we are interested in computing the probability 
distribution $P_\omega(V)$ defined as
\begin{align}
\label{eq:pw1}
P_\omega(V) &=
\sum_{N=0}^\infty\frac{1}{N!}
\int\dots\int\,P\bigl\{X_1\dots X_N\bigl\lvert\Phi_0(V)\bigr\} \\
& \qquad \times \omega(\vx_1)\dots\omega(\vx_N)\, e^{{\cal W}_0(V)} \nonumber \;,
\end{align}
where $\omega(\vx)$ is a position-dependent weight and the multiplicative factor of 
$1/N!$ reflects the fact that the objects are identical.
Details of the calculation can be found in Appendix \S\ref{app:CIC}. 
In short, substituting the explicit expression of $P\{X_1\dots X_N|\Phi_0(V)\}$, which 
involves products of the conditional correlation functions ${\cal W}_N$, collecting 
the terms of same order in $\bar{n}$ shows that the series expansion Eq.(\ref{eq:pw1}) 
nicely re-sums into the compact expression
\begin{equation}
\label{eq:pomega}
P_\omega(V) = e^{{\cal W}_\omega(V)}-e^{{\cal W}_0(V)} \;.
\end{equation}
The probability $P_0=\exp({\cal W}_0)$ of an empty cell is subtracted because it does
not carry any weight.
Furthermore, in analogy with (minus) the conditional void correlation ${\cal W}_0(V)$, 
we have defined
\begin{align}
\label{eq:womega}
{\cal W}_\omega(V) &= \sum_{k=1}^\infty\frac{(-\bar{n})^k}{k!}
\int_V\!\!\dd^3\!\vx_1\dots\int_V\!\!\dd^3\!\vx_k\,\xi_k(\vx_1,\dots,\vx_k) \nonumber \\
& \qquad \times \bigl(1-\omega(\vx_1)\bigr)\dots\bigl(1-\omega(\vx_k)\bigr) \;.
\end{align}
Note the similarity of this expression with the partition function $Z[J]$ introduced by 
\cite{szapudi/szalay:1993}. Eq.(\ref{eq:womega}) indeed is their $Z[J]$ with a source 
term $J(\vx) = \omega(\vx) - 1$. 

\subsection{Application to the UV ionising background}\label{sec:app_UV}

The characterisation of fluctuations in the ionising background generated by clustered 
sources provides an interesting application for our weighted probability distribution 
$P_\omega(V)$.

Namely, let $\{\vx_k\}$, $k=1,\dots,N$, be the comoving positions of $N$ quasars 
distributed inside a cell of volume $V\propto R^3$ at redshift $z$. Each of them emits 
ionising radiation, so that the angle-averaged specific intensity of ionising photons 
(in units of $\junit$) at the centre of the cell is
\begin{equation}
\label{eq:intensity}
J_k(\vx_k)=(1+z)^2 \frac{L_k}{(4\pi r_k)^2} e^{-r_k/r_0} \;.
\end{equation}
$r_k=|\vx_k|$ is the modulus of the separation vector, $L_k$ is the quasar luminosity 
(in $\lunit$) and $r_0$ is the attenuation length of ionising photons in the intergalactic 
medium. We will hereafter ignore the multiplicative factor of $(1+z)^2$ and quote specific 
intensities relative to their mean. This factor should, of course, be re-introduced in order 
to compute the absolute photoionisation rate $\Gamma$ etc.

The probability to have an angle-averaged specific intensity $J$ at the centre of the cell 
is obtained upon summing over all the configurations subject to the constraint $\sum_k J_k=J$. 
In other words, each configuration of $N$ sources contributes a factor of
\begin{multline}
\int\!\!\dd\alpha_1\dots \dd\alpha_N\, \phi(\alpha_1)\dots \phi(\alpha_N) \\
\times P\bigl\{X_1\dots X_N\Phi_0(V)\bigr\} \\ 
\times \delta_D(J_1+\dots+J_N-J)
\label{eq:1config}
\end{multline}
to the total probability. The measure $\phi(\alpha)d\alpha$ with $\alpha=L/L^\star$ is the 
probability density for the quasar luminosity, $L^\star=L^\star(z)$ is a characteristic, 
usually redshift-dependent luminosity and the specific intensity $J_k$ now reads 
$J_k=\alpha_k L^\star \exp(-r_k/r_0)/(4\pi r_k)^2$. 
We will henceforth assume that the reduced correlations $\bar{\xi}_k$ do not depend on 
$\alpha$, yet our results can be straightforwardly extended to include a dependence of 
clustering on $\alpha$.

Substituting the Laplace representation of the Dirac delta in Eq.(\ref{eq:1config}),
\begin{equation}
\delta_D(J_1+\dots+J_N-J) = \frac{1}{2\pi i}\int_{-i\infty}^{+i\infty}\!\!
\dd s\,e^{s(J-J_1-\dots-J_N)}\;,
\end{equation}
integrating the variables $\vx_1$, ... , $\vx_N$ and summing over $N\geq 0$, we find that the 
probability $P(J)$ to have a total specific intensity $J$ at the centre of a cell of volume 
$V$ is exactly given by Eq.(\ref{eq:pw1}) with a weight
\begin{equation}
\label{eq:weightJ}
\omega(\vx_k)=\Theta_H(R-|\vx_k|)
\int_{\alpha_{\rm min}}^{\alpha_{\rm max}}\!\!\dd\alpha_k\,\phi(\alpha_k)\,
e^{-s J_k(\vx_k)}
\end{equation}
assigned to each object. The Heaviside function $\Theta_H(R-|\vx|)$ delimits the cell boundaries, 
while the lower and upper limits of the integral are $\alpha_{\rm min}=L_{\rm min}/L^\star$, 
$\alpha_{\rm max}=L_{\rm max}/L^\star$. Finally,  $s$ is the variable conjugate to $J$. 
Therefore, $P(J)$ takes the compact form
\begin{equation}
\label{eq:laplace}
P(J) = \frac{1}{2\pi i}
\int_{-i\infty}^{+i\infty}\!\!\dd s\,e^{sJ+{\cal W}_\omega(V)} \;,
\end{equation}
where the weight $\omega$ is given by Eq.(\ref{eq:weightJ}).
The contribution from the void conditional probability can be ignored since it is independent of 
$s$ and, therefore, only contributes at $J=0$ (empty cells do not generate any radiation).
In other words, $P(J)$ truly is the distribution of intensity conditioned on the cell being not
empty. It is, of course, normalised to unity. 

As we will see shortly, the Laplace transform yields a more intuitive description than the Fourier 
transform. In practice however, the Fourier representation of the Dirac delta turns out to be more
convenient for the numerical evaluation of $P(J)$~:
\begin{equation}
\label{eq:fourier}
P(J) = \frac{1}{2 \pi} \int_{-\infty}^{+\infty} \!\!\dd s\, e^{-isJ+{\cal W}_\omega(V)}\;,
\end{equation}
with the weight given by 
\begin{equation}
\omega({\vx}_k) = \Theta_H(R-|\vx_k|)
\int_{\alpha_{\rm min}}^{\alpha_{\rm max}}\!\!\dd \alpha_k\, \phi(\alpha_k)\, e^{is J_k({\vx}_k)}
\end{equation}
The numerical implementation will be discussed in more detail in \S\ref{sec:numerics}.

\begin{figure}
\center
\resizebox{0.45\textwidth}{!}{\includegraphics{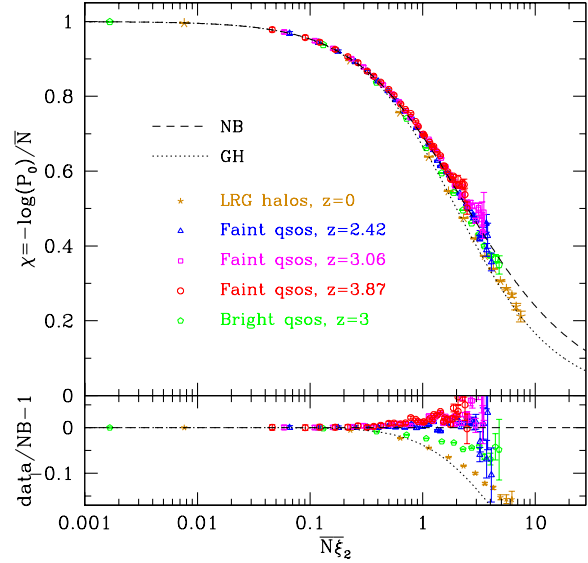}}
\caption{{\it Top panel}~: Void scaling function $\chi$ as a function of average clustering 
strength $\bar{N}\bar{\xi}_2$.
(Orange) stars represent $\chi$ for $z=0$ haloes. (Blue) triangles, (magenta) squares and (red) 
circles indicate the scaling for the mock quasars of the ``faint'' sample at $z=2.4$, 3 and 
3.9, respectively. (Green) pentagons are $\chi$ for the ``bright'' sample at $z=3$ (see text). 
The dashed and dotted curves indicate the void scaling in the Negative Binomial (NB) and 
Geometric Hierarchical (GH) models, respectively. 
{\it Bottom panel}~: Fractional deviation from the NB void scaling (colour online).}
\label{fig:void}
\end{figure}

\subsection{Specialisation to hierarchical models}

The evaluation of Eqs.(\ref{eq:laplace}) or (\ref{eq:fourier}) is not a trivial task since it 
requires knowledge of all reduced correlation functions $\xi_k$ of the 
sources. Interestingly however, $P(J)$ can be easily computed when the clustering of sources 
follow the hierarchical ansatz. In this case, all the information about source clustering is 
contained in the 2-point correlation function and the void scaling function.

\subsubsection{Hierarchical scaling and random dilutions}
\label{sec:scaling}

In the hierarchical approximation, volume-averaged correlation functions are of the form 
$\bar{\xi}_k = S_k \bar{\xi}_2^{k-1}$ where the coefficients $S_k$ (which are ratios of
connected moments) are generally scale-independent, and converge towards $S_k=k^{k-2}$ in
the rare halo limit \citep{bernardeau/schaeffer:1999}.
Hence, we can recast the logarithm of the void probability into the series \citep{fry:1986a}
\begin{align}
\label{eq:hscaling}
{\cal W}_0(V) &= -\bar{N}
\sum_{k=1}^\infty \frac{(-1)^{k-1}}{k!} S_k\left(\bar{N}\bar{\xi}_2\right)^{k-1} \\
&\equiv -\bar{N}\chi(\bar{N}\bar{\xi}_2) \nonumber \;.
\end{align}
Consequently, the void scaling function 
\begin{equation}
\chi = - \frac{\ln\bigl(P_0\bigr)}{\bar{N}} = -\frac{{\cal W}_0(V)}{\bar{N}}
\end{equation}
depends on the distance $r$ through $\bar{\xi}_2(r)$ only. Note that we recover 
$\chi\equiv 1$ for a pure Poisson distribution $P_0=e^{-\bar{N}}$, whereas $0<\chi < 1$ 
holds for any clustered distribution.

Even though observational data 
\citep{bouchet/strauss/etal:1993,gaztanaga:1994,croton/gaztanaga/eta:2004,ross/brunner/myers:2006}
and numerical simulations \citep{fry/colombi/etal:2011} indicate that the hierarchical
amplitudes $S_k$ of the galaxy distribution depend on scale, the simulated and observed void 
probabilities appear to obey the hierarchical scaling Eq.(\ref{eq:hscaling}).
As shown by \cite{fry/colombi:2013}, this can be explained by the halo model if the distribution
of host haloes follows the hierarchical pattern. Moreover, one should expect that different 
populations of tracers are described by different void scaling relations.

Several analytic formulae have been proposed for the void scaling function 
\citep[see][for a discussion]{fry:1986a}.
Comparison with N-body simulations indicate that the geometric hierarchical 
\citep[GH, e.g.][]{carruthers/shih:1983} 
and negative binomial \citep[NB, e.g.][]{hamilton:1988} 
models are good approximation for galaxies (sub-haloes) and haloes extracted from N-body 
simulations, respectively \citep{fry/colombi:2013}. 
The corresponding functional form of $\chi$ is
\begin{align}
\chi(\bar{N}\bar{\xi}_2) &=
\frac{\ln\!\left(1+\bar{N}\bar{\xi}_2\right)}{\bar{N}\bar{\xi}_2} \qquad \mbox{(NB)} \\
\chi(\bar{N}\bar{\xi}_2) &=
\frac{1}{1+\frac{1}{2}\bar{N}\bar{\xi}_2} ~\qquad\quad \mbox{(GH)} 
\end{align}
Clustering becomes significant in the regime $\bar{N}\bar{\xi}_2\gtrsim 1$, i.e. high number 
densities and/or large correlation length. 

Random dilutions of a point distribution will affect the average number density $\bar{N}$ but not 
the correlation functions $\bar{\xi}_k$ 
\citep{peebles:1980,lahav/saslaw:1992,sheth:1996}. However, while the void scaling functions 
of the parent and diluted sample generally differ, some distributions preserve their functional
form. As shown in \cite{sheth:1996}, this is the case of the NB model. 
This can easily be seen upon rewriting the generating functional $g(\lambda)=\sum_N P_N \lambda^N$ 
as
\begin{equation}
g(\lambda) = \bigl[1 + \bar{N}\bar{\xi}_2(1-\lambda)\bigr]^{-1/\bar{\xi}_2}  \;.
\end{equation}
Since a random dilution is equivalent to the transformation $\lambda\to p\lambda+q$, where 
$p<1$ is the dilution factor and $q=1-p$ \citep{lahav/saslaw:1992}, we find 
\begin{equation}
g(p\lambda+q) = \bigl[1 + p\bar{N}\bar{\xi}_2(1-\lambda)\bigr]^{-1/\bar{\xi}_2} \;,
\end{equation}
which shows that the diluted distribution follows the NB scaling with a number density $p\bar{N}$ 
\citep{sheth:1996}. 

\subsubsection{Quasar void scaling function}
\label{sec:quasars}

For {\small HeII}-reionization discussed in Sec.\S\ref{sec:results}, quasars are the relevant 
ionising sources. 

In order to ascertain whether the void scaling function of quasars also follows the hierarchical 
scaling without going into a detailed modelling of their distribution, we use the synthetic quasar 
catalogues of \cite{croton:2009} extracted from the {\small MILLENNIUM} simulation
\citep{springel/white/etal:2005}. These catalogues were constructed by abundance matching under
the assumption that quasars populate both parent haloes and sub-haloes above the minimum resolved 
halo mass, i.e. $M_{\rm min}\sim 10^{11}\hmsun$. Quasars are thus randomly sub-sampling (sub)halo 
centres of mass $M>M_{\rm min}$ with a dilution factor $p$ equal to their duty cycle $f=t_Q/t_H$. 
Here, $t_Q$ and $t_H$ are the typical quasar lifetime and the Hubble time at redshift $z$, 
respectively \citep{martini/weinberg:2001,haiman/hui:2001}. 
We adopt a duty cycle of $f\approx 0.037$, which leads to a quasar number density of 
$\bar{n}\approx 6.0\times 10^{-4}\hhhmpc$ at $z=3$.
We consider three samples at $z=2.42$, 3.06 and 3.87, which we refer to as the ``faint'' 
quasars since they include (sub)haloes down to a relatively small mass. 

Since the quasar demographics are relatively uncertain, we generate an additional mock catalogue. 
We assume that quasars populate only parent haloes above the minimum mass, although small-scale 
clustering measurements indicate that a halo may host more than one shining quasar 
simultaneously \citep{hennawi/strauss/etal:2006,myers/richards/etal:2008,padmanabhan/white/etal:2009}.
This should be a reasonable assumption at high redshift and for separations $r\gtrsim 1\hmpc$ 
larger than the typical halo scale \citep{conroy/white:2013}.
We use dark matter haloes extracted from N-body simulations evolving 1024$^3$ particles in periodic 
boxes of size 1500$\hmpc$ \citep[for details about the simulations, see][]{biagetti/chan/etal:2014}. 
We sample all haloes above the minimum resolved halo mass. i.e. $M_{\rm min}=5\times 10^{12}\hmsun$
and $p=1$. 
We will refer to this sample as the ``bright'' quasars since they only trace massive haloes. 
We focus on the snapshot at $z=3$. The corresponding quasar number density is
$\bar{n}\approx 5.1\times 10^{-4}\hhhmpc$, close to that of the ``faint'' sample.

Following \cite{fry/colombi:2013}, we compute the void probability $P_0$, the mean $\bar{N}$ and 
the variance in excess of Poisson 
$\bar{N}^2\bar{\xi}_2=\bigl\langle N^2\bigr\rangle-\bar{N}^2-\bar{N}$ 
from non-overlapping cells with radius in the range $R=1-40\hmpc$. The uncertainty on $\chi$ is 
calculated following the prescription of \cite{colombi/bouchet/schaeffer:1995}.
While Fig.\ref{fig:void} clearly shows that, for the ``faint'' samples, the data closely 
follows the NB scaling, there is compelling evidence that the void scaling function of the ``bright''
sample lies between the NB and GH scalings, despite the lack of data for $\bar{N}\bar{\xi}_2$ 
much larger than unity. 
Notwithstanding, our measurements strongly suggest that the void scaling function of quasars also 
follows the hierarchical pattern, but the scaling may depend on the details of the quasars
demographics. We will henceforth assume that it is well represented by the NB model around $z=3$. 
We thus expect random dilutions of the quasars population to preserve the NB scaling.

As a consistency check, we have also computed $\chi$ for the low redshift haloes that host 
luminous red galaxies (LRGs), i.e. the $z=0$ haloes with $M>5\times 10^{12}\hmsun$. We have found 
that their void scaling function is better represented by the GH model,in agreement with the 
findings of \cite{fry/colombi:2013}. 
In all cases, the various measurements converge towards the Poisson value $\chi\equiv 1$ in the 
limit $\bar{N}\bar{\xi}_2\ll 1$ (i.e. infinitesimal cell radius), as expected.

\subsection{UVB fluctuations in hierarchical models}
\label{sec:hierarchUVB}

\begin{figure}
\center
\resizebox{0.45\textwidth}{!}{\includegraphics{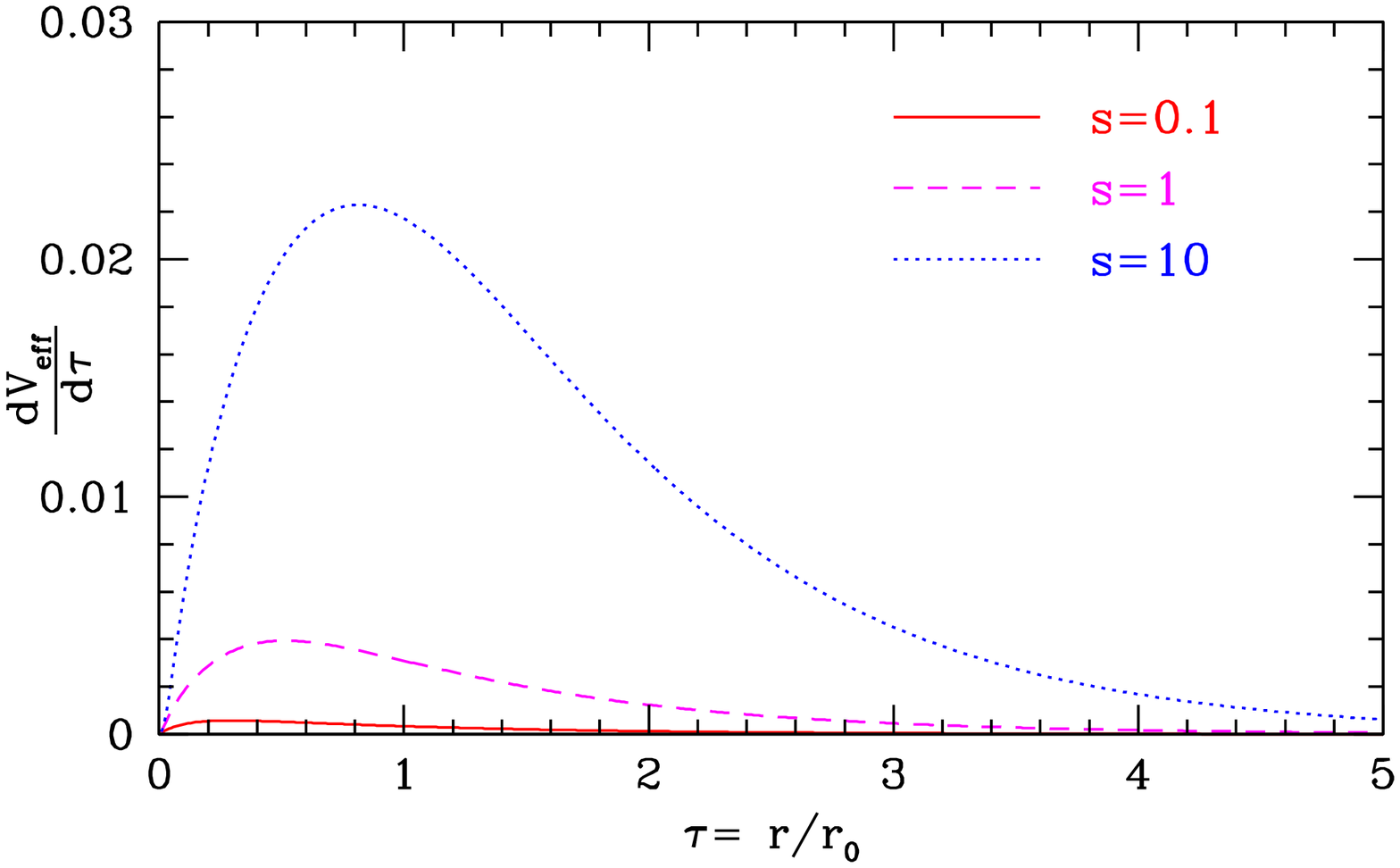}}
\resizebox{0.45\textwidth}{!}{\includegraphics{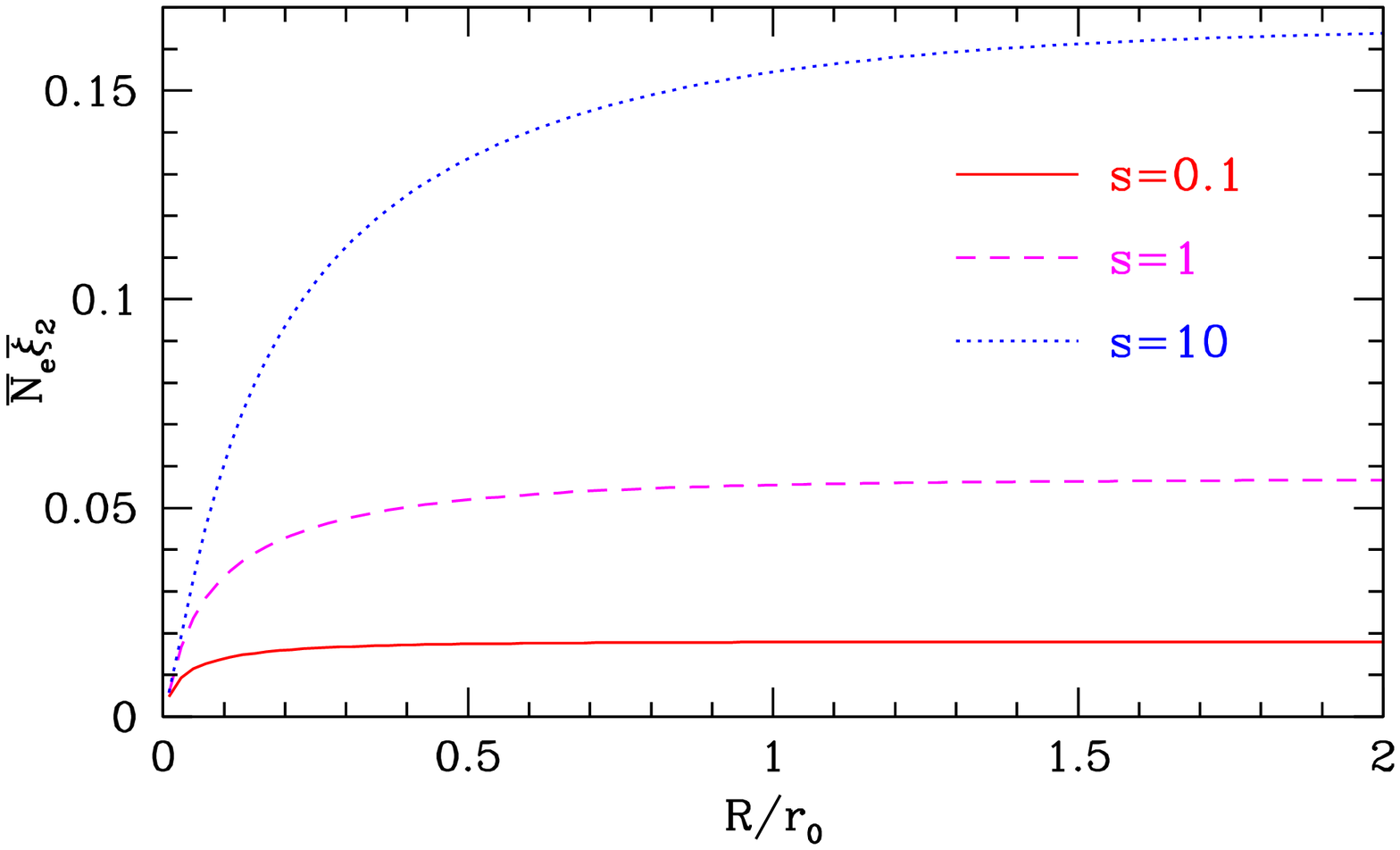}}
\caption{{\it Top panel}~: Differential effective volume $\dd V_e/\dd\tau(s,\tau)$ (in unit of 
$3\bar{N}_0/\bar{n}$) as a function of the optical depth $\tau=r/r_0$.
{\it Bottom panel}~: Average clustering strength $\bar{N}_e\bar{\xi}_2$ (in unit of  
$3\tau_\xi^\gamma\bar{N}_0$) as a function of the dimensionless cell radius $R/r_0$.
Results are shown for three different values of $s$ assuming the two-power-law form 
Eq.(\ref{eq:qlf}) for the quasar luminosity function, and a power-law correlation function 
with logarithmic slope $\gamma=1.9$ (colour online).}
\label{fig:Veff}
\end{figure}

The hierarchical ansatz holds regardless the shape of the window function that defines the cell 
of volume $V$ as long as it decays sufficiently rapidly to zero for large $\vx$. 
This suggests that we could also assume some sort of hierarchical scaling for the weighted void 
probability ${\cal W}_\omega(V)$ since the window function is effectively
\begin{equation}
\Theta_H(|\vx|-R)\,\bigl(1-\omega(\vx)\bigr)\;.
\end{equation}
The term $1-\omega(\vx)$ will always suppress the contribution of regions with $|\vx|\gg 1$, 
even when the cell size $R$ is very large. For concreteness, let us have a closer look at the 
effective volume 
\begin{equation}
\label{eq:Veff1}
V_{\rm e}(s,V) \equiv \int\!\!\dd^3\!\vx\,\bigl(1-\omega(\vx)\bigr)W_T(\vx,V) \;,
\end{equation}
which is the relevant quantity in our calculation of UV background fluctuations.
Following \cite{meiksin/white:2003}, we introduce the normalised specific intensity $j=J/J^\star$, 
with $J^\star=L^\star/(4\pi r_0)^2$, the optical depth $\tau=r/r_0$ at a distance $r$ from the 
source and the average number of ionising sources $\bar{N}_0=(4\pi/3)r_0^3\bar{n}$ within an
attenuation volume. The effective volume becomes 
\begin{align}
\label{eq:Veff}
V_{\rm e}(s,V) &\equiv \int_0^{R/r_0}\!\!\dd\tau\,\frac{\dd V_e}{\dd\tau}(s,\tau) \\
&= 3\left(\frac{\bar{N}_0}{\bar{n}}\right)\int_0^{R/r_0}\!\!\dd\tau\,\tau^2
\nonumber \\ & \qquad \times 
\int_{\alpha_{\rm min}}^{\alpha_{\rm max}}\!\!\dd\alpha\,\phi(\alpha) 
\left(1-e^{-s\alpha\tau^{-2}e^{-\tau}}\right)
\nonumber \;.
\end{align}
where $R$ is the radius of the tophat filter. The extra factor of $J^\star$ as been absorbed 
into the redefinition $s\to sJ^\star$, such that $s$ and $j$ are conjugate variables.
The top panel of Fig.\ref{fig:Veff} displays the behaviour 
of $\dd V_e/\dd\tau$ as a function of optical depth for a few choices of $s$. For illustration 
purposes, $\dd V_e/\dd\tau$ is plotted in unit of $3\bar{N}_0/\bar{n}$ assuming the usual double 
power-law form for the quasar luminosity function (see Eq.~\ref{eq:qlf}).
$\dd V_e/\dd\tau$ reaches a global maximum and decays as $\exp(-\tau)$ in the limit $\tau\gg 1$, 
suggesting indeed that the hierarchical approximation holds also when the sources are weighted by
their contribution to the specific intensity at $\vx=0$.

Therefore, under the assumption that the hierarchical ansatz discussed above also applies for the
weighted tophat window $\Theta_H(|\vx|-R)\bigl(1-\omega(\vx)\bigr)$, the probability distribution 
$P(j)=P(J)J^\star$ for the normalised specific intensity $j$ is
\begin{equation}
\label{eq:PJ1}
P(j) = \frac{1}{2\pi i}
\int_{-i\infty}^{+i\infty}\!\!\dd s\,e^{sj+{\cal W}_\omega(V)} \;,
\end{equation}
with a weighted, conditional void probability given by 
\begin{equation}
\label{eq:WV1}
{\cal W}_\omega(V)=\sum_{k=1}^\infty\frac{\left(-\bar{N}_{\rm e}\right)^k}{k!}
\bar{\xi}_k(V_{\rm e})
\equiv -\bar{N}_{\rm e}\,\chi\bigl[\bar{N}_{\rm e}\bar{\xi}_2(V_{\rm e})\bigr]\;.
\end{equation}
$\bar{N}_e=\bar{n} V_e$ is the mean number count in the effective volume $V_e$ and 
\begin{align}
\label{eq:NX1}
\bar{N}_e\bar{\xi}_2 &\equiv \left(\frac{\bar{n}}{V_e}\right)
\int\!\!\dd^3\!\vx_1\int\!\!\dd^3\!\vx_2\,\xi_2(\vx_1,\vx_2) \bigl(1-\omega(\vx_1)\bigr) \\
&\qquad \times\bigl(1-\omega(\vx_2)\bigr)
\Theta_H(|\vx_1|-R)\Theta_H(|\vx_2|-R) \nonumber 
\end{align}
is the corresponding integrated clustering strength.

Consider the large bubble limit $R\gg r_\xi$, where $r_\xi$ is the characteristic clustering
length of the sources, so that the volume-average 2-point correlation function is approximately 
$\bar{\xi}_2 \sim V^{-1}\int_Vd^3\vx\, \xi_2(r)$. In this regime, 
\begin{align}
\left(\bar{N}_e\bar{\xi}_2\right)\!\!(s) &\approx 3\bar{N}_0
\int_{\alpha_{\rm min}}^{\alpha_{\rm max}}\!\!\dd\alpha\,\phi(\alpha) \\
& \qquad \times 
\int_0^{R/r_0}\!\!\dd\tau\,\tau^2 \xi(\tau)\left(1-e^{-s\alpha\tau^{-2}e^{-\tau}}\right)
\nonumber \;.
\end{align}
For a power-law 2-point correlation $\xi_2(r)=(r/r_\xi)^{-\gamma}$, $\bar{N}_e\bar{\xi}_2$ 
saturates in the limit $R/r_0\gg 1$ as can be seen in the bottom panel of Fig.~\ref{fig:Veff}, 
where $\bar{N}_e\bar{\xi}_2$ is shown in unit of $3\tau_\xi^\gamma\bar{N}_0$
Furthermore, $\bar{N}_e\bar{\xi}_2$ increases with $s\sim 1/j$.
We thus naively expect that clustering effects shall be large for $j\ll 1$, but relatively 
small for $j\gg 1$ since the product $\bar{N}_e\bar{\xi}_2$ saturates rapidly when $s\ll 1$. 

Eqs. (\ref{eq:PJ1}) -- (\ref{eq:NX1}) are the main result of this Section. We will now explore
the behaviour of $P(j)$ in the regime $j\ll 1$ and $j\gg 1$ before discussing its practical
(numerical) implementation.

\section{Asymptotics and numerics} 
\label{sec:assymp}

\subsection{Asymptotic expressions}
\label{sec:assym}

Inverse Laplace transforms are notoriously difficult to perform. Nevertheless, we can use the 
saddle point approximation to derive closed analytic expressions for the low- and high-intensity 
tails. Our analysis proceeds along the lines of 
\cite{bernardeau/kofman:1995,colombi/bernardeau/etal:1997,valageas:2002,valageas/munshi:2004,
bernardeau/codis/pichon:2013}. 
As will be shown shortly, there is a critical intensity $j_c$ such that, for $j\ll j_c$, the 
saddle point dominates the contribution to the integral whereas, for $j\gg j_c$, it is the 
critical point that controls the asymptotic behaviour. For illustration purposes, we will only
consider the limit $V\to\infty$, but the same conclusions hold for finite bubble radii. Details
of the calculation can be found in Appendix \S\ref{app:asymptotics}.

\subsubsection{Random sources}

We begin with the simpler case of randomly-distributed sources. The weighted conditional void 
probability reduces to ${\cal W}_\omega(V)\equiv -\bar{n}V_e(s,V)$. Integrating over the optical
depth by parts in Eq.(\ref{eq:Veff}) and subsequently taking the limit $V\to\infty$, we 
arrive at \citep{meiksin/white:2003}
\begin{equation}
V_e(s,V\to\infty) = s \left(\frac{\bar{N}_0}{\bar{n}}\right)
\int_{\alpha_{\rm min}}^{\alpha_{\rm max}}\!\!\dd\alpha\,\alpha\phi(\alpha)\,h(-s \alpha)\;,
\end{equation}
where the function $h(x)$ is
\begin{equation}
\label{eq:gx}
h(x) \equiv \int_0^\infty\!\!\dd\tau\exp\!\left(x \tau^{-2}e^{-\tau}\right)e^{-\tau}
\left(2+\tau\right)\;.
\end{equation}
Performing the inversion $s\to -s$ through the origin in Eq.(\ref{eq:PJ1})
~\footnote{The purpose of this inversion is to deal with 
Legendre transforms of convex rather than concave functions, see below.}, 
the probability distribution for the normalised intensity $j$ takes the form
\begin{equation}
\label{eq:PJ2}
P(j) = \frac{1}{2\pi i}\int_{-i\infty}^{+i\infty}\!\!\dd z\, e^{-zj+G(z)}\;,
\end{equation}
with
\begin{equation}
\label{eq:Gz}
G(z) = -\bar{N}_e(s=-z) = 
z \bar{N}_0 \int_{\alpha_{\rm min}}^{\alpha_{\rm max}}\!\!\dd\alpha\,\alpha\phi(\alpha)\,
h(z \alpha)\;.
\end{equation}
The function $G(z)$, where $z=x+iy$ is the complex variable, is the continuation of 
${\cal W}_\omega(V)$ over the complex plane. 
$G$ is analytic everywhere except along the positive real axis $x>0$ where it is not 
defined, and it has a branch point at $z=0$ where $G(0)=0$.
On the negative real axis, $G(x)$ is a convex, monotonically increasing function of 
$x$, i.e. $G(x)\leq 0$ for $x\leq 0$.

\begin{figure}
\center
\resizebox{0.45\textwidth}{!}{\includegraphics{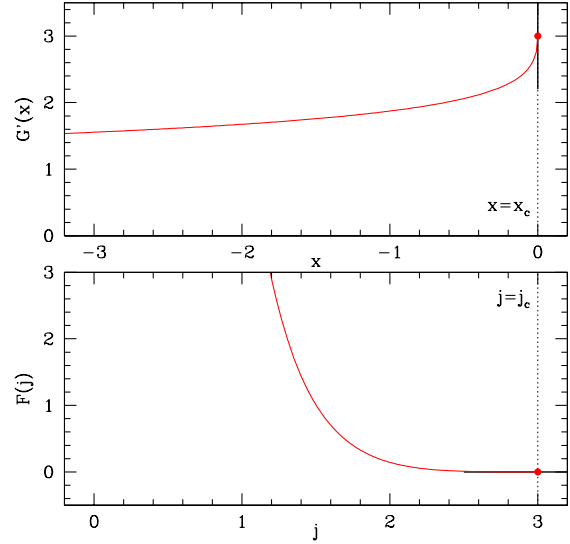}}
\caption{Graph of the first derivative of $G(x)$ (top panel) and its Legendre transform $F(j)$ 
(bottom panel) near the critical point $(x,j)=(x_c,j_c)$. Both $G(x)$ and $F(j)$ have the same 
convexity.}
\label{fig:legendre}
\end{figure}

The argument of the exponential in Eq.(\ref{eq:PJ2}) admits a saddle point along the 
negative real axis of the complex plane which is amenable to a stationary phase (or 
steepest descent) calculation if
\begin{align}
\frac{\partial}{\partial x}\bigl(-xj+G(x)\bigr) &=0 \\
\frac{\partial^2}{\partial x^2}\bigl(-xj+G(x)\bigr) &> 0 
\end{align}
The first condition implies $j=G'(x)$.
As shown in the top panel of Fig.\ref{fig:legendre}, it can be satisfied for $j\leq j_c$
solely, where the critical intensity  
$j_c=3\bar{N}_0\bigl\langle\alpha\bigr\rangle\equiv \bigl\langle j\bigr\rangle$ is also
the mean specific intensity \citep{meiksin/white:2003}.
The second condition guarantees that the real part $G(x)$ goes through a local maximum 
when $z$ varies perpendicular to the real axis. This must be true since $G(x)$ is 
convex over the whole negative real axis. 

Consider $j<j_c$ and let $(x_s,0)$ (with $x_s<0$) be the coordinate of the corresponding saddle 
point in the complex plane. We can expand $-zj+G(z)$ along the path $z=x_s+iy$ for small 
$|y|\ll 1$~: 
\begin{equation}
-jz+G(z) \approx -j x_s + G(x_s) - \frac{1}{2} G''(x_s)\, y^2 + \dots
\end{equation} 
where a prime denotes a derivative w.r.t. the variable $x$, and we have used the fact that the 
real and imaginary part of $G$ are harmonic. At this point, it is convenient to introduce an
auxiliary function $F(j)$ defined as the Legendre transform of $G(x_s)$, i.e.
\begin{equation}
\label{eq:legendre}
F(j) + G(x_s) = j x_s
\end{equation}
with $j=G'(x_s)$ and $x_s=F'(j)$. Differentiating Eq.(\ref{eq:legendre}) w.r.t. to $j$ or $x_s$, 
we recover the well-known relation $G''(x_s) = F''(j)^{-1}$. Hence, we can also write
$-zj+G(z)\approx -F(j) - y^2 / \bigl(2 F''(j)\bigr)$. Taking the constant piece out of the inverse
Laplace transform and performing the remaining Gaussian integral over $y$, we obtain the usual
formula
\begin{equation}
\label{eq:saddle}
P(j) \approx \sqrt{\frac{F''(j)}{2\pi}}\, e^{-F(j)}\;.
\end{equation}
Finally, taking the limit $(x,j)\to (-\infty,0)$ and using the Legendre transform to solve for 
$x(j)$, we arrive at (see Appendix \S\ref{app:low})
\begin{equation}
\label{eq:saddle1}
P\bigl[\ln(j/j_c)\bigr] = 
-\sqrt{\frac{3\bar{N}_0}{2\pi}} \bigl(\ln(j/j_c)\bigr)^{-1}
e^{\bar{N}_0\ln^3(j/j_c)}\;,
\end{equation}
where $P(\ln j)=j P(j)$. Even though this expression is only valid in the limit of small 
intensities, we shall expect a sharp cutoff when $j\lesssim j_c$. This is clearly seen in 
Fig. \ref{fig:pj1}.

When $j>j_c$, the contour in the complex plane is pushed along the real positive axis, and wraps
around the critical point $z_c=0$ where the second derivative $G''(x)$ becomes singular. In this
case, the trick consists in expanding $F(j)$ around $j_c$ (see the bottom panel of Fig.
\ref{fig:legendre}) rather than $G(x)$ around $x_c=0$, and 
exploiting the fact that both functions are Legendre transforms of each other to derive an 
expression for $G(x)$ valid around $x_c$. We retain only the dominant singular contribution to $G$
to obtain the leading contribution to $P(j)$. The argument of the exponential admits the series 
expansion (see Appendix \S\ref{app:high})
\begin{equation}
\label{eq:argzc}
-zj+G(z) = -\left(j - j_c\right) z - \frac{2}{3}\sqrt{\frac{2}{f_3}}\, z^{3/2} + \dots
\end{equation}
where $f_3\equiv F^{(3)}(j_c)$ is a negative real number. Sub-leading contributions scale as $z^2$, 
$z^{5/2}$ etc. On performing the integral in the complex plane, we arrive at
\begin{equation}
\label{eq:pjasym1}
P(j) \approx \frac{1}{\sqrt{2\pi}}\left(j-j_c\right)^{-5/2} {\rm Im}\bigl(-f_3^{-1/2}\bigr) \;.
\end{equation}
Lastly, we compute $f_3$ by taking advantage of relations between the derivatives of the Legendre
transforms $F$ and $G$. 
We find $f_3=-(2/9\pi)\bigl(\bar{N}_0\bigl\langle\alpha^{3/2}\bigr\rangle\bigr)^{-2}$, so that
\begin{equation}
\label{eq:pjasym2}
P(j) \approx \frac{3}{2} \bar{N}_0\bigl\langle\alpha^{3/2}\bigr\rangle\left(j-j_c\right)^{-5/2} \;.
\end{equation}
This scaling agrees with that found by \cite{meiksin/white:2003} except for an additional, 
multiplicative  factor of 2.

\subsubsection{Clustered sources}

As seen in Sec. \S\ref{sec:theory}, source clustering can be taken into account upon assuming 
that the conditional void correlation is of the form Eq.(\ref{eq:hscaling}). In this case, we
can perform an analysis similar to the random case if we define
\begin{align}
\label{eq:pjclust}
P(j) &= \frac{1}{2\pi i}\int_{-i\infty}^{+i\infty}\!\!\dd z\, e^{-zj+{\cal G}(z)} \\
{\cal G}(z) &= G(z) \chi(z) \;,
\end{align}
where $G(z)$ is given by Eq.(\ref{eq:Gz}) and $\chi(z)=\chi[\bar{N}_e\bar{\xi}_2(s=-z)]$.
Hence, it is sufficient to study the behaviour of the void scaling function $\chi(z)$ in order 
to ascertain the impact of source clustering on the low- and high-intensity tail of the 
distribution. 
We clearly have $\chi(z)\to 1$ when we approach the critical point $z_c=0$. 
Furthermore, on the negative real axis, $\chi(x)$ is a monotonically increasing function of 
$x$ that vanishes in the limit $x\to -\infty$.

For any choice of $j<j_c$, ${\cal G}(x)$ also exhibits a saddle-point on the negative real 
axis. However, since $0<\chi(x)<1$ is monotonically increasing, the saddle-point position $(x_s,0)$ 
in the complex plane is closer to the origin than for randomly-distributed sources. 
As a result, $-F(j)=-x_s j +G(x_s)$ is less negative. Therefore, we also expect a cutoff at low 
intensities, but it should occur at relatively smaller values of $j$. 
For a power-law correlation $\xi_2(r)=(r/r_\xi)^{-2}$, a quick computation yields
\begin{equation}
\label{eq:pjlowclust}
P(j) \sim 
\exp\left[-\frac{1}{3\tau_\xi^2}\ln^2\!\left(\frac{3\tau_\xi^2 j}{2\langle\alpha\rangle}\right)\right]\;,
\end{equation}
where $\tau_\xi\equiv r_\xi/r_0$ is the source correlation length $r_\xi$ in unit of the attenuation
length. Clearly, a slight increase in $r_\xi$ will result in a large amplification of the probability
$P(j)$ owing to the exponential factor. Moreover, the dependence on $\ln^2(j)$ rather than
$\ln^3(j)$ suggests that the cutoff is not as sharp as in the random case.

Source clustering also affects the amplitude of the distribution in the high-intensity regime. 
For the power-law correlation $\xi_2(r)=(r/r_\xi)^{-2}$, we find
\begin{equation}
\label{eq:pjhighclust}
P(j) \approx \frac{3}{2} \left(1+A\bar{N}_0\tau_\xi^2\right) 
\bar{N}_0 \bigl\langle\alpha^{3/2}\bigr\rangle\left(j-j_c\right)^{-5/2}\;,
\end{equation}
where the coefficient $A$ is proportional to moments of the source luminosity function. 
A simple approximation to the average clustering strength $(\bar{N}_e\bar{\xi}_2)(s=-z)$ around 
$z=0$ leads to $A= (9/4)\langle\sqrt{\alpha}\rangle\langle\alpha\rangle/\langle\alpha^{3/2}\rangle$.

\subsection{The mean intensity}

The mean specific intensity $\langle j\rangle$ does not change if source clustering 
is turned on, regardless the value of $R$. 
To see this, we write $\langle j\rangle = \int\! dj\,j P(j)$, substitute Eq.(\ref{eq:pjclust}) 
and integrate $j e^{-zj}$ by part. We are thus left with
\begin{equation}
\langle j \rangle = \frac{1}{2\pi i}\int_{\gamma-i\infty}^{\gamma+i\infty}\!\! \dd z \,
\frac{e^{{\cal G}(z)}}{z^2} = \text{Res}\left(z^{-2}e^{{\cal G}(z)},z=0\right) \;.
\end{equation}
Since ${\cal G}(z)\approx 3\bar{N}_0\langle\alpha\rangle(1-e^{-R/r_0}) z +{\cal O}(z^{3/2})$ in 
the limit $z\to 0$, the residue is always $3\bar{N}_0\langle\alpha\rangle(1-e^{-R/r_0})$, i.e. 
the mean intensity for a bubble radius $R$ \citep{meiksin/white:2003}. 
This demonstrates our assertion.

\subsection{Numerical implementation} 
\label{sec:numerics}

In what follows, we use the Fourier transform to evaluate the probability distribution $P(j)$ 
numerically. Symmetry considerations show that $P(j)$ is equal to the real part of 
Eq.(\ref{eq:fourier}),
\begin{align}
P(j) &= \frac{1}{\pi} \int_0^\infty\!\! \dd s \, \cos\bigl( -s j + \rm{Im} \ {\cal G}(-s)\bigr)  \\
      &\qquad   \times \exp\bigl({\rm Re}\ {\cal G}(-s)\bigr) \nonumber \;.
\end{align}
Even though the integrand is highly oscillatory at large $s$, its envelop is damped exponentially
in this regime as Re$\,{\cal G}(-s) \to -\infty$ in the limit $s\to\infty$. Therefore, the 
integral converges very well even when $j$ is significantly larger than $j_c$. In practice, we
sample the real and imaginary part of ${\cal W}_\omega(V)$ evenly in $\log(s)$ with ${\cal O}(10)$
points per decade from $s=10^{-5}$ to $s=10^3$. We use the {\small VEGAS} Monte-Carlo algorithm 
\citep{lepage:1978} to evaluate the 5-dimensional integrated clustering strength Eq.(\ref{eq:NX1}).  
We subsequently perform the integral over $s$ using a Gauss-Konrod quadrature. Note that
\begin{equation}
\int_{-1}^{+1}\!\!\dd\mu\bigl(\tau_1^2+\tau_2^2-2\tau_1\tau_2\mu\bigr)^{-\gamma/2}
=\frac{\left(\tau_1+\tau_2\right)^{2-\gamma}-\bigl\lvert\tau_1-\tau_2\bigr\lvert^{2-\gamma}}
{\left(2-\gamma\right)\tau_1\tau_2}\;,
\end{equation}
which can be used to reduce the dimensionality of the integral Eq.(\ref{eq:NX1}) in the case of a
power-law correlation function $\xi_2(r)=(r/r_\xi)^{-\gamma}$. 

To test the accuracy of our numerical results, especially for values of $j\ll j_c$ where the 
impact of source clustering is expected to be most significant, we also compute $P(j)$ from the 
inverse Laplace transform Eq.(\ref{eq:laplace}). In practice, we begin by inverting the relation 
$j={\cal G}'(z)$ to locate the saddle point $z=x_s$ on the negative real axis, and then build a 
path of constant phase in the complex plane upon imposing the condition
\begin{equation}
\delta\bigl(-z j+{\cal G}(z)\bigr) \in \mathbb{R}
\end{equation}
for each infinitesimal step $\delta z$ \citep[see e.g.][for similar constructions]
{colombi/bernardeau/etal:1997,valageas:2002,bernardeau/codis/pichon:2013}. We use the adaptive,
multi-dimensional algorithm {\small CUBATURE} \citep{berntsen/espelid/genz:1991} to compute 
${\cal G}(z)$ and its first derivative. The computation of ${\cal G}'(z)$ slows down the Laplace
transform considerably relative to the Fourier transform. 

\section{Results}
\label{sec:results}

In this Section, we discuss the effect of source clustering on the probability density $P(j).$
For the sake of illustration, we consider the effect of quasar clustering on fluctuations in the 
{\small HeII}-ionising background at the end of helium reionization ($z\sim 3$). We shall 
make a few simplifying assumptions here as our goal is not to model the intergalactic medium 
in detail, but merely obtain a reasonable estimate of the effect. We defer a more detailed 
study to future work.

\subsection{Model inputs: quasars and the IGM}

We need to determine three quantities in order to calculate the probability $P(j)$ of the 
{\small HeII}-ionising radiation: the quasar luminosity function $\Phi(L,z)$, the quasar 2-point 
correlation function $\xi_2(r)$ and the attenuation length $r_0$ of the HeII ionising photons.

A finite cell radius $R$ implies that only a finite number of sources can illuminate a random 
field point. This happens prior to the completion of helium reionization, when the ionised bubbles 
around quasars are surrounded by neutral gas which absorbs the radiation emitted by sources outside 
the local region \citep[see e.g.][]{furlanetto:2008}. Since we consider the end of helium 
reionization, $R$ is formally infinite. In practice, we shall take $R=1000$ much larger than the
values of $r_0$ considered.

\subsubsection{Quasar luminosity function}

We parametrise the bolometric quasar luminosity function (QLF), which we define as the differential 
comoving number density of quasars with bolometric luminosity $L$ and redshift $z$, with the standard 
double power-law form 
\citep[e.g.][]{boyle/shanks/peterson:1988,boyle/griffiths/etal:1993,pei:1995,croom/smith/etal:2004},
\begin{equation}
\label{eq:qlf}
\Phi(L,z) = 
\frac{\Phi_\star(z)/L_\star(z)}
{(L/ \ L_\star(z))^{\beta_1(z)}+(L/L_\star(z))^{\beta_2(z)}}\;,
\end{equation}
where $\Phi_\star$ is a normalisation, $\beta_1(z)$ and $\beta_2(z)$ are the faint- and bright-end 
slopes of the distribution, respectively, and the characteristic luminosity $L^\star(z)$ marks the 
break from a shallow to a steep slope. Eq.(\ref{eq:qlf}) furnishes a good representation of the 
observations if one allows $\Phi_\star$, $\beta_1$, $\beta_2$ and $L_\star$ to vary with redshift. 
We use the best-fit values inferred by \cite{hopkins/etal:2007} for the quasar bolometric QLF at 
$z=3$, i.e.
\begin{align}
\Phi_\star &= 2.56 \times 10^{-6} \ {\rm Mpc}^{-3} \nonumber \\
L_\star &= 10^{13.17} \ {\rm L}_{\sun} \nonumber \\
 \beta_1 &= 1.395  \nonumber \\
\beta_2 &= 3.10
\end{align}  
The normalised, dimensionless quasar number density $\phi(\alpha)$ is constructed from $\Phi(L,z)$ 
from the relation \citep{meiksin/white:2003}
\begin{equation}
\phi(\alpha)=\frac{\Phi(\alpha L^\star)L^\star}{\int_{L_{\rm min}}^{L_{\rm max}}\!\!dL\,\Phi(L)} \;.
\end{equation}
At the bright end, the number density of quasars diminishes so rapidly that the exact value of
$L_{\rm max}$ has little impact on the results. However, the integral is quite sensitive to 
$L_{\rm min}$ owing to the much shallower faint-end slope. In what follows, we will assume
$L_{\rm max}=5\times 10^{14}\ {\rm L}_\odot$ and $L_{\rm min}=10^{10}\ {\rm L}_\odot$. This yields
a total quasar number density of $\bar{n}\approx 10^{-4}$ Mpc$^{-3}$, while the abundance of 
$L>L_\star$ quasars is only $\approx 8\times 10^{-7}$ Mpc$^{-3}$.

At this point, we should in principle convert the bolometric quasar luminosity $L$ into an ionising
intensity at the frequencies of interest (i.e. $h\nu\geq h\nu_\text{\tiny HeII}=54.4$eV) assuming, 
for instance, that the quasar spectral energy distribution follows the broken power-law template of 
\cite{madau/haardt/rees:1999}. 
However, since our main objective is to illustrate
the applicability of our count-in-cell approach, we will ignore this conversion and only present 
distributions for the normalised intensity $j=J/J_\star$. In doing so, we do not take into account
the scatter in the far-UV spectral index \citep{telfer/zheng/etal:2002,desjacques/nusser/sheth:2007}, 
Nevertheless, this should have a negligible impact on the intensity distribution $P(j)$
\citep[see Fig.1 of][]{furlanetto:2008}.

\subsubsection{Quasar clustering}

The real-space 2-point correlation function of quasars is often fitted to a power-law of the form 
$\xi_2(r)=(r/r_\xi)^{-\gamma}$. A number of studies have explored the clustering of high-redshift 
quasars, but their clustering amplitude is still a matter of debate. 
Early estimates based on the incidence of close quasar pairs set lower limits to the correlation 
length of $r_\xi \gtrsim 15 - 20$ Mpc
\citep{stephens/schneider/etal:1997,kundic:1997,schneider/fan/etal:2000,djorgovski/stern/etal:2003}. 
In a more recent analysis based on a sample of $4462$ quasars in the redshift range $2.9\leq z \leq 5.4$,
\cite{shen/strauss/etal:2007} obtained $r_\xi \sim 21$ Mpc assuming a power-law slope $\gamma \sim 2$, 
with a strong indication that the high-redshift quasars with $z\geq 3.5$ are substantially more 
clustered ($r_\xi \sim 35$ Mpc).
\cite{francke/gawiser/etal:2008} found a similar, albeit smaller value of $r_\xi\sim 14$ Mpc from a
measurement of the cross-correlation between Lyman Break Galaxies (LBGs) and quasars in the redshift
range $2.7<z<3.8$.
In what follows, we will fix the power-law slope to $\gamma=2.1$, but let the correlation length vary 
generously around the fiducial value of $r_\xi=15$ Mpc. Furthermore, we shall assume that the quasar
void scaling function follows hierarchical clustering, i.e. $\chi=\chi(\bar{N}\bar{\xi}_2)$, and is 
well represented by the Negative Binomial model discussed above. Note that $\chi$ needs not be 
universal. 
Our approximation would indeed work even if $\chi$ depends on redshift because the attenuation length 
$r_0$ is considerably smaller than the Hubble time $t_H$. However, it is crucial that $\chi$ be a 
function of the integrated clustering strength $\bar{N}\bar{\xi}_2$ only. 

These analyses also suggest that quasar clustering strongly depends on luminosity at high redshift,
in agreement with various theoretical predictions 
\citep{porciani/magliocchetti/norberg:2004,hopkins/etal:2007,croton:2009}. Even though our procedure 
ignores this possibility, we stress that the model of \cite{croton:2009} predicts a linear bias 
$b_1\sim 6-9$ (assuming $\sigma_8=0.9$) for the $z=3$ quasars shining at the characteristic luminosity 
$L_\star$, consistent with our choice of $r_\xi=15$ Mpc for the fiducial quasar correlation length.

\subsubsection{Attenuation length and cell size}

The (comoving) attenuation length of {\small HeII}-ionising photons is a crucial ingredient of our model. 
Following \cite{furlanetto:2008,dixon/furlanetto:2009}, we shall ignore variations in the sight line 
opacity and any frequency-dependence in order to characterise this attenuation through a single number
$r_0$. Estimations based on the incidence of Lyman-limit systems \citep{bolton/haehnelt/etal:2006} or
the propagation of ionising photons around individual quasars \citep{furlanetto/oh:2008} indicate that
the average attenuation length at $z=3$ is $r_0\sim 30-40$ Mpc, while the more sophisticated treatment 
of \cite{davies/furlanetto:2014} yields a somewhat larger value, $r_0\sim 60$ Mpc.
To be conservative, we will consider a couple of attenuation lengths in addition to the fiducial value of 
35 Mpc so as to brackets the aforementioned estimates.

\subsection{Intensity distribution in a fully ionised IGM}

\begin{figure}
\center
\resizebox{0.45\textwidth}{!}{\includegraphics{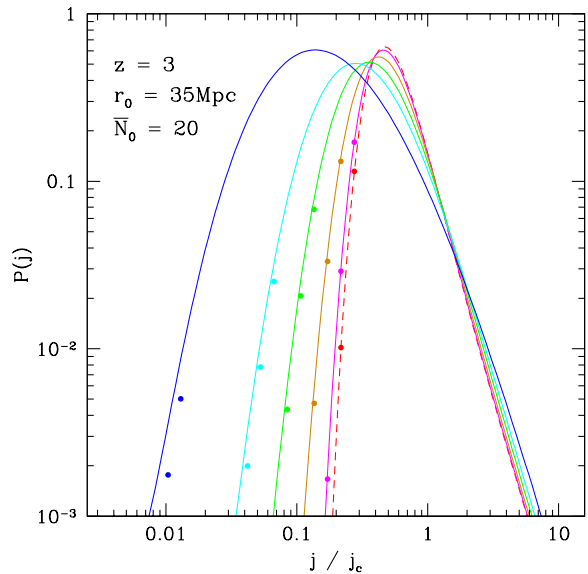}}
\caption{The probability distribution $P(j)$ of the {\small HeII}-ionising intensity (in unit of 
$j_c$) in the reionized IGM. The attenuation length is $r_0=35$ Mpc and the QLF is the standard 
double power-law Eq.(\ref{eq:qlf}). The dashed curve is for randomly distributed quasars, whereas 
the solid curves assume a power-law correlation with correlation length $r_\xi=5$, 10, 15, 20 and 
30 Mpc from narrowest to widest, respectively. 
The data points have been obtained upon applying the saddle point approximation to the Laplace 
transform Eq.(\ref{eq:laplace}), and are in good agreement with the various curves, which have 
all been computed using the Fourier transform Eq.(\ref{eq:fourier}) (colour online).}
\label{fig:pj1}
\end{figure}

Fig.\ref{fig:pj1} illustrates the effect of quasar clustering on the distribution of 
{\small HeII}-ionising
intensity for an attenuation length $r_0=35$ Mpc. All the distributions have been computed using 
the Fourier transform Eq.(\ref{eq:fourier}). For comparison, the data points have been obtained 
from the Laplace transform Eq.(\ref{eq:laplace}) using the saddle point approximation described 
in \S\ref{sec:assym}.
The good agreement between the two methods demonstrates that our numerical evaluation of $P(j)$
is robust. The dashed line is for randomly distributed quasars, whereas the solid curves show 
$P(j)$ for a quasar correlation length in the range $5<r_\xi<30$ Mpc (increasing from the narrowest 
to the widest distribution). Clustering widens the distribution at small $j$ essentially because 
source correlations substantially increase the probability of finding regions devoid of quasars.
The effect becomes significant when $r_\xi\gtrsim 15$ Mpc for the attenuation length adopted here.
At high intensity, the various distributions converge towards the scaling $P(j) \sim j^{-5/2}$.
The amplitude increases with clustering strength, in agreement with our asymptotic expectation 
Eq.(\ref{eq:pjhighclust}). 
The probability to have an intensity $j\gtrsim 3 j_c$ is $\sim 17$\% (resp. 80\%) larger for 
$r_\xi=15$ Mpc (resp. $r_\xi=30$ Mpc) relative to randomly distributed quasars. 
This fairly weak enhancement is consistent with  a value of $A\approx 0.05$ in 
Eq.(\ref{eq:pjhighclust}) much smaller than 
$A= (9/4)\langle\sqrt{\alpha}\rangle\langle\alpha\rangle/\langle\alpha^{3/2}\rangle\approx 0.3$
expected for the QLF adopted here.

\begin{figure}
\center
\resizebox{0.45\textwidth}{!}{\includegraphics{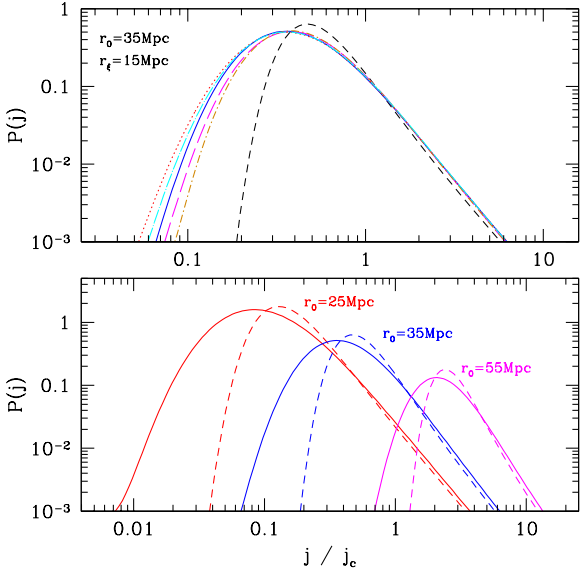}}
\caption{{\it Top panel}~: Effect of changing the behaviour of the quasar correlation function 
on the distribution $P(j)$. The short-dashed (black) line represents the Poisson case.
The solid (blue) curve is our fiducial model, the dotted (red) curve 
was obtained using the GH rather than the NB void scaling function, the long-dashed (magenta) curve 
has $\xi_2=0$ for $r<1$ Mpc while the dotted-short dashed (orange) assumes $\xi_2=0$ outside the range 
$1<r<150$ Mpc. Finally, the dotted-long dashed (cyan) curve assumes a power-law slope $\gamma=1.9$
rather than 2.1.
The correlation and attenuation lengths are $r_\xi=15$ Mpc and $r_0=35$ Mpc, respectively.
{\it Bottom panel}~: $P(j)$ for 3 different attenuation lengths. Results are shown for
randomly distributed (dashed curves) and clustered sources with $r_\xi=15$ Mpc (solid curves).}
\label{fig:pj2}
\end{figure}

The high-intensity scaling $P(j)\sim j^{-5/2}$ reflects the behaviour of the nearest neighbour 
probability density, 
\begin{equation}
\label{eq:neighbour}
H_1(r)\dd r = -\frac{\partial P_0}{\partial r}\dd r = \frac{\partial(\bar{N}\chi)}{\partial r}
e^{-\bar{N}\chi}\dd r\;.
\end{equation}
Consider indeed that all the quasars shine with a luminosity $L=L_\star$. Ignoring the attenuation 
of ionising photons, the optical depth scales as $\tau=(J_\star/J)^{1/2}=j^{-1/2}$. For a random
distribution,
\begin{equation}
H_1(\tau)\dd\tau = 3\bar{N}_0 \tau^2 e^{-\bar{N}_0\tau^3}\,\dd\tau
\end{equation}
and is, of course, normalised to unity: $\int\!\!\dd\tau\, H_1(\tau) = 1$. 
On inserting $\tau=j^{-1/2}$ into this expression, we derive a probability density
\begin{equation}
P(j)\dd j = \frac{3}{2}\left(\frac{\bar{N}_0}{j^{5/2}}\right) e^{-\bar{N}_0/ j^{3/2}} \,\dd j
\end{equation}
for the ionising intensity. 
Finally, replacing $\bar{N}_0$ by $\bar{N}_0 \bigl\langle\alpha^{3/2}\bigr\rangle$ yields the scaling 
Eq.(\ref{eq:pjasym2}). This scaling persists in the clustered case since, in the limit $\tau\ll 1$, 
the nearest neighbour distribution is insensitive to the amplitude of clustering.
Most importantly however, the amplitude increases with the clustering strength as discussed above,
presumably because finding the second-nearest neighbour close to the first one is more likely.

We have thus far assumed that the quasar 2-point correlation follows a power-law at all separations,
even though we expect quasars to be anti-correlated at very large scales. Furthermore, if quasars
populate distinct haloes, then we should also expect anti-correlation at separations $r\lesssim 1$ Mpc
smaller than the typical halo size. In order to gauge the importance of these effects, the top panel 
of Fig.\ref{fig:pj2} displays the distribution $P(j)$ for the fiducial power-law scaling, yet assuming
$\xi_2(r)=0$ at short separations $r<1$ Mpc (long-dashed curve), as well as outside the range 
$1<r<150$ Mpc (dotted-dashed curve).
In this case, we have checked that setting $\xi_2=-0.001$ or $-0.01$ for $r>150$ Mpc does not change 
$P(j)$ appreciably.
We also show the impact of changing the void scaling function from the fiducial NB scaling to the GH
model (dotted curve), and raising the power-law slope from $\gamma=1.9$ to 2.1 (dotted-long dashed 
curve). 
The Poisson case is also overlaid on this figure (short-dashed curve) for comparative purposes.
Overall, the low-intensity tail is quite sensitive to variations in the default
assumptions, with up to an order of magnitude difference in the probability already at $j=0.1 j_c$.  
By contrast, the high-intensity tail is barely affected as it is dominated by the nearest neighbour.

\begin{table}
\caption{Variance $\langle\Delta j^2\rangle$ of intensity fluctuations relative to the Poisson 
result. Both the quasar clustering length $r_\xi$ and the attenuation length $r_0$ are in units 
of (comoving) Mpc.}
\vspace{1mm}
\begin{center}
\begin{tabular}{cccccc} 
\hline
 & $r_\xi=5$ & $r_\xi=10$ & $r_\xi=15$ & $r_\xi=20$ & $r_\xi=30$ \\
\hline\hline
$r_0=25$ & 1.02  & 1.08 & 1.19 & 1.35 & 1.56 \\
\hline
$r_0=35$ & 1.03  & 1.10 & 1.23 & 1.41 & 1.95 \\
\hline
$r_0=55$ & 1.05  & 1.15 & 1.32 & 1.56 & 2.25 \\
\hline
\end{tabular}
\end{center}
\label{table1}
\end{table}

The impact of clustering relative to Poisson fluctuations should diminish as the number 
density $\bar{N}_0$ of sources in an attenuation volume decreases. This is indeed the case, as we 
will see shortly. At low intensities however, the opposite happens. To understand this, consider
the GH scaling for simplicity. For $j\ll 1$ (i.e. $s\gg 1$), the weighted, conditional void 
probability ${\cal W}_\omega$ given by Eq.(\ref{eq:WV1}) scales like
$-\bar{N}_e\chi(\bar{N}_e\bar{\xi}_2)\sim -\bar{\xi}_2^{-1} \sim -\tau_\xi^{-\gamma}$ when 
$\bar{N}_e\bar{\xi}_2\to\infty$. In other words, ${\cal W}_\omega$ increasingly deviates from the 
Poisson results $\sim - r_0^3$, and a larger clustering length further enhances the suppression, in
agreement with our asymptotic expression Eq.(\ref{eq:pjlowclust}).
This is clearly seen in Fig.\ref{fig:pj2}, where the intensity distributions for randomly-distributed 
and clustered sources are compared for three different values of the attenuation length 
$r_0=25$, 35 and 55 Mpc. The number of quasars in an attenuation volume is $\bar{N}_0\sim 7$, 20 and 
76, respectively. A constant clustering length $r_\xi=15$ Mpc is assumed for all the solid curves. 
Note again the enhancement of $P(j)$ at large intensities, which is consistent with 
Eq.(\ref{eq:pjhighclust}) (i.e. the effect increases with $\bar{N}_0 \tau_\xi^2\sim r_0r_\xi^2$) 
provided that $A\approx 0.05$. 
 
To quantify the impact of source clustering on $P(j)$, we have measured the variance of intensity
fluctuations, $\langle\Delta j^2\rangle=\langle j^2\rangle-\langle j\rangle^2$, relative to the 
Poisson case for a range of values of $r_0$ and $r_\xi$. 
Results are summarised in Table \ref{table1}. All the models assume a power-law slope $\gamma=2.1$. 
As expected, the deviation increases with $r_0$ or, equivalently, with decreasing Poisson noise. 
At fixed $r_0$, it echoes the rise in the amplitude of the $j^{-5/2}$ tail with increasing 
correlation length $r_\xi$. 

\subsection{Environmental dependence of $P(j)$}
\label{sec:env}

We have thus far focused on the distribution $P(j)$ for random field points.
Source clustering increases the probability for intensities $j\ll j_c$ because regions devoid of 
quasars are significantly more abundant. Therefore, we may expect that $P(j)$ depends on whether we
sit in a high or low density region. 

\subsubsection{Spherical collapse considerations}

To ascertain the magnitude of this environmental dependence, we restrict the set of field points to
those located at the centre of spheres of volume $V\propto R^3$ with fractional density $\delta$. 
The conditional void probability function acquires a dependence on $\delta$,
\begin{align}
{\cal W}_0(V|\delta) &=\sum_{k=0}^\infty \frac{(-\bar{n})^k}{k!}
\int_V\!\!\dd^3\!\vx_1\dots\int_V\!\!\dd^3\!\vx_k\,\xi_k(\vx_1,\dots,\vx_k|\delta) 
\nonumber \\
&\equiv \sum_{k=1}^\infty\frac{(-\bar{N})^k}{k!}\bar{\xi}_k(V|\delta) \;.
\end{align}
As before, ${\cal W}_0(V|\delta)$ generates all the count probabilities subject to the condition that 
the cell fractional density is $\delta$. In particular, since $\xi_1(\vx|\delta)$ is now different 
from unity, the average number density of sources in those cells, 
\begin{equation}
\left\langle N|\delta\right\rangle V^{-1} =\bar{n} V^{-1}\int_V\!\!\dd^3\!\vx\;\xi_1(\vx|\delta)\;,
\end{equation}
is a decreasing (increasing) function of $V$ if $\delta>0$ ($\delta < 0$) such that
$\left\langle N|\delta\right\rangle \to \bar{n}V$ in the limit of large cell volume. In other
words, $\xi_1(\vx|\delta)\equiv \xi_1(r|\delta)$ is the average source density profile around a 
given overdensity $\delta$.

\begin{figure}
\center
\resizebox{0.45\textwidth}{!}{\includegraphics{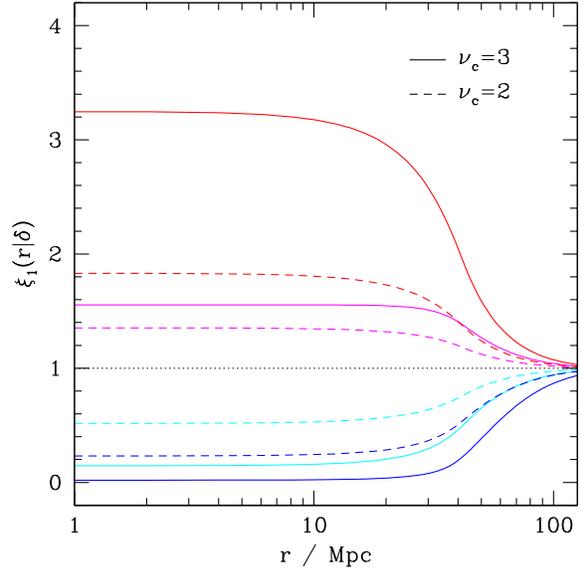}}
\caption{Relative abundance of sources $\xi_1(r|\delta)=\bar{n}(r|\delta)/\bar{n}$ around overdense
and underdense regions as a function of comoving distance $r$. Results are shown for a large-scale 
environment density $\delta=-2\sigma$, $-1\sigma$, $+1\sigma$ and $+2\sigma$ at $z=3$ (curves from 
bottom to top), where $\sigma$ is the rms variance of the evolved density field on comoving scale 
$R=28.5$ Mpc. Solid and dashed lines were obtained assuming $\nu_c=3$ and 2, respectively.}
\label{fig:profile}
\end{figure}

To estimate $\xi_1(r|\delta)$, we use the spherical collapse model, which establishes a connection 
between the evolved region and the initial seed perturbation \citep{gunn/gott:1972,peebles:1980}. 
Namely, the initial size $R_0$ and overdensity $\delta_0$ are related to $R$ and $\delta$ through 
\citep{bernardeau:1994,mo/white:1996,sheth:1998}
\begin{equation}
\delta_0= \delta_c \Bigl[1-\bigl(1+\delta\bigr)^{-1/\delta_c}\Bigr]\;,\qquad
1+\delta = \left(\frac{R_0}{R}\right)^3 \;.
\end{equation}
Here, $\delta_0$ is the initial density linearly extrapolated to the redshift under consideration, 
so it can take values less than $-1$. These relations can be used
to estimate the initial profile $\xi_1(s|\delta_0)=\bar{n}(s|\delta_0)/\bar{n}$ as a function of 
Lagrangian separation $s$.
Let $R_1$ be the characteristic Lagrangian radius of the peaks that collapse into the haloes 
hosting quasars.
In the peak-background split approach \citep{kaiser:1984}, density fluctuations in the environment 
locally modulate the peak number density. Taking into account the non-zero correlation between
$R_1$ and $R_0$, the initial profile is
\begin{align}
\xi_1(s|\delta_0) &= \frac{{\cal N}(\nu_c|\nu_0,s)}{{\cal N}(\nu_c)} \\
&= \exp\left[-\frac{\epsilon^2(s)\bigl(\nu_c^2+\nu_0^2\bigr)-2\epsilon(s)\nu_c\nu_0}
{2\bigl(1-\epsilon^2(s)\bigr)}\right] \nonumber \;,
\end{align}
where ${\cal N}$ is a Normal distribution,
$\nu_c=\delta_c/\sigma_1$ is the peak height, $\nu_0=\delta_0/\sigma_0$ is the significance
of the initial large-scale perturbation, and $\epsilon(s)=\sigma_\times^2(s)/(\sigma_0\sigma_1)$ 
is the cross-correlation between the short- and long-wavelength modes. 
Here, $\sigma_0$ and $\sigma_1$ are the rms variance of density fluctuations smoothed on scale 
$R_0$ and $R_1$, respectively, and
\begin{equation}
\sigma_\times^2(s)=\frac{1}{2\pi^2}\int_0^\infty\!\!dk\,k^2\,P(k)\,W_T(kR_0) W_T(kR_1)
j_0(ks)
\end{equation}
is a cross-correlation involving one filter of size $R_0$ and the other of size $R_1$. 
Evolving $\bar{n}(s|\delta_0)$ requires in principle knowledge of the average, initial density 
profile as a function of $s$. For simplicity however, we assume that $\xi_1(s|\delta_0)$
evolves in a self-similar way, and convert Lagrangian to Eulerian scales according to 
$r = (1+\delta)^{-1/3}s$. Therefore, we compute $\xi_1(r|\delta)$ as
\begin{equation}
\xi_1(r|\delta) = \xi_1\bigl(s(r)|\delta_0\bigr) \;.
\end{equation}
Fig.\ref{fig:profile} displays several profiles obtained for a large-scale environment density 
$\delta=-2\sigma$, $-1\sigma$, $+1\sigma$ and $+2\sigma$ (curves from bottom to top), where 
$\sigma$ is the rms variance of the $z=3$ density field smoothed on comoving scale $R=28.5$ Mpc. 
The effect sensitively depends on the choice of $\nu_c$. 
Dashed and solid curves assume a peak height $\nu_c=2$ and 3 (obtained upon setting $R_1=1.1$
and $3$ Mpc), which correspond to linear halo 
biases $b_1\sim 1 + (\nu_c/\sigma_1)\approx 3.4$ and 6.4, respectively. Low density regions with
$\delta=-2\sigma$ hardly contain virialized, $\nu_c=3$ haloes.

Furthermore, all the higher-order correlations $\bar{\xi}_k(V|\delta)$ are also affected by the 
environmental constraint. Their $\delta$-dependence could also be worked out using the spherical
collapse model. However, since we can only speculate about whether the void scaling function 
$\chi(r|\delta)$ still satisfies the hierarchical scaling, we will present results assuming 
$\bar{\xi}_k(V|\delta)=0$ for $k\geq 3$. Note that the sources are nonetheless clustered to some
extent since their number density increases (decreases) in overdense (underdense) regions 
as exemplified in Fig.\ref{fig:profile} 

The top panel of Fig.\ref{fig:pj3} displays the resulting conditional distribution $P(J|\delta)$ 
given a large-scale environment density $\delta=-2\sigma$, 0 and $+2\sigma$ (dashed, dotted and 
solid curves, respectively). 
The corresponding average intensity is $\langle J\rangle\approx 0.96$, 2.70 and 5.84 in unit of 
$J_{35}^\star\equiv J^\star(r_0=35\text{Mpc})$, as is the abscissa of Fig.\ref{fig:pj3}.
We have assumed $\nu_c=3$ as above to compute $\xi_1(r|\delta)$, and a fixed attenuation length 
$r_0=35$ Mpc regardless the value of $\delta$. Even though the differences in $P(j)$ are quite 
significant, they should be regarded as an upper bound since we have considered relatively rare, 
2$\sigma$ fluctuations traced by highly biased sources. 

\subsubsection{Sensitivity to the clustering of absorption systems}

Clearly, the attenuation length must vary spatially since it is mainly determined by the number 
density of absorption systems. While the absorption systems with low {\small HI} column densities 
(i.e. $\log(N_{\rm\tiny HI})<17.2\cmm$) are distributed relatively uniformly, both the Lyman Limit
Systems (LLS;  $17.2<\log(N_{\rm\tiny HI})<20.3\cmm$) 
-- which correspond to metal line ({\small Mg II}, {\small C IV}) systems -- 
and Damped Ly$\alpha$ Absorbers (DLA; $\log(N_{\rm\tiny HI})>20.3\cmm$) 
-- which trace gas-rich galaxies at high redshift -- 
are expected to be noticeably clustered, though likely not as much as quasars. 
For instance, the recent analysis of \cite{font-ribera/miralda-escude/etal:2012} finds $b_1\sim 2.2$ 
for DLAs in the redshift range $2<z<3.5$.
Clearly, strong absorption systems will be overabundant (underabundant) in regions with 
$\delta>0$ ($\delta<0$). Hence, we might expect a relatively shorter (longer) attenuation length 
when the ionising radiation field is seen from the centre of an overdense (underdense) region.

\begin{figure}
\center
\resizebox{0.45\textwidth}{!}{\includegraphics{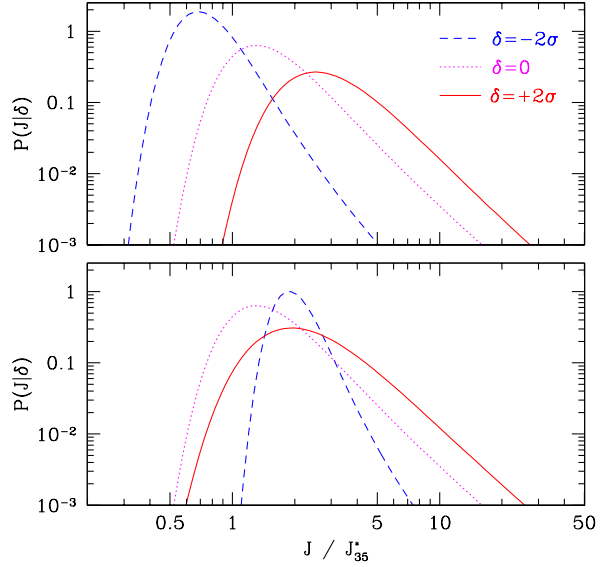}}
\caption{Distribution $P(j)$ as a function of the large-scale environment, characterised by the 
fractional density $\delta$ on comoving scale $R=20\hmpc$. In the top panel, a constant attenuation 
length $r_0=35$ Mpc is assumed regardless of the large-scale density whereas, in the bottom panel, 
$r_0$ is allowed to vary with the local environment density (see text). Note that the specific 
intensity $J$ is in unit of $J^\star_{35}\equiv J^\star(r_0=35\,{\rm Mpc})$.}
\label{fig:pj3}
\end{figure}

The clustering length of absorption systems generally depends on their column density. However, 
owing to the scarcity of observational constraints, we simply assume that the absorption systems 
trace the $\nu_c=2$ peaks discussed above and set the local attenuation length to 
$r_0(\delta) = r_0\, \xi_1(0|\delta)^{-1/3}$ in the computation of $P(j)$. 
This scaling reflects the fact that $r_0\propto \bar{n}_\text{abs}^{-1/3}$, where $\bar{n}_\text{abs}$
is the number density of LLS and DLAs.
The resulting attenuation length is $\sim 57$ Mpc and $\sim 29$ Mpc for the regions with large-scale 
density $\delta=-2\sigma$ and $+2\sigma$, respectively. 
The corresponding intensity distributions are shown in the bottom panel of Fig.\ref{fig:pj3}. 
Unsurprisingly, our spatially-varying prescription for $r_0(\delta)$ reduces differences between 
the distributions obtained for low and high density regions. Still, the average intensity in
$\delta=+2\sigma$ regions remains about twice as large (5.06) as that of random field points.
Although a detailed account of the clustering of absorption systems around the sources will be
essential to quantify this effect precisely, it is clear that variations in the mean intensity 
should not exceed a few, even for relatively pronounced overdense or underdense regions.

\section{Discussion}
\label{sec:discussion}

To our knowledge, the only study which has thus far addressed the impact of quasar clustering on 
the post-reionization distribution $P(j)$ is the semi-numerical treatment of 
\cite{dixon/furlanetto/mesinger:2014}, in which dark matter haloes are generated upon applying the 
excursion set approach to realisations of the linear density field in periodic boxes of size 
$L=250\hmpc$.
Overall, our results are consistent with theirs: clustering widens the intensity distribution and, 
thus, enhances the probability for $j\ll \langle j\rangle$ and $j\gg\langle j\rangle$. 
Regarding the magnitude of the effect, the fact that their distributions are nearly identical 
regardless of whether the sources randomly sample haloes or are randomly distributed suggests that 
the correlation length $r_\xi$ of their synthetic quasars is fairly small. 
Furthermore, their fiducial attenuation length is $r_0=60$ Mpc, about twice as large as ours. 
The effects shown in Fig.\ref{fig:pj1} would appear smaller, had we adopted the same value of $r_0$.
Finally, their simulated distributions exhibit a very sharp cutoff at low intensities, presumably 
because their simulation box is too small to contain a representative sample of those underdense 
regions responsible for the low-intensity tail.

We have derived asymptotic expressions to further check the validity of our numerical implementation. 
While our asymptotic scaling is consistent with the impact of source clustering as inferred from 
the numerical evaluation of $P(j)$, there is a mismatch at high intensities between the 
analytic and the numerical prediction of the power-law tail. 
Namely, the numerical evaluation of Eq.(\ref{eq:fourier}) shows that the effect of source
clustering is $\sim 6$ times smaller than predicted by the asymptotic expectation Eq.(\ref{eq:highasym}).
We have not been able to understand the origin of this discrepancy, and a more rigorous analysis
is beyond the scope of this paper. However, we have numerically checked that, for a fixed $r_0$, 
the mean intensities of the distributions with $0\leq r_\xi\leq 30$ Mpc agree with each at the 
1.5 percent level (large differences would offset the distribution). 
Therefore, there is no systematic offset along the abscissa.
We thus believe that the rise of the high-$j$ amplitude with $r_\xi$ is real, rather than the 
manifestation of a numerical error.

The main drawback of our method is the absence of a treatment for the small-scale structure of 
the IGM, radiative transfer effects etc. 
\citep[see e.g.][]{maselli/ferrara:2005,tittley/meiksin:2007} 
Notwithstanding, it has the advantage to be very fast -- 
generate a distribution $P(j)$ takes ${\cal O}(20)$ minutes on a standard workstation -- and, 
thus, allows us to explore a wide range of quasar properties and demographics. 
In the present study, the model inputs are the observed quasar luminosity function and 2-point 
correlation, but one could instead use predictions based on a halo occupation distribution (HOD).
The large scatter in the observed correlation length of high redshift quasars may reflect, at 
least partly, a luminosity-dependence of quasar clustering. Our approach can be
extended to account for this dependence: the source correlation functions could in principle 
depend on both $r$ and $\alpha$, and the behaviour of the void scaling function could generally 
be a function of $\alpha$.
Further improvements include a frequency-dependent attenuation length (to account for the longer 
mean free path of hard photons) and a better modelling of the clustering of absorption systems. 
Clearly, such analytic approaches will never surpass detailed (and computationally expensive) 
cosmological hydrodynamical simulations with radiative transfer, but they can furnish useful 
insights into the effect of discrete, clustered sources and absorbers on the physical state of 
the high-redshift IGM.

Finally, one should bear in mind the caveat that the weighted void probability follows the 
hierarchical ansatz (see Sec.\S\ref{sec:hierarchUVB}). 
We have shown that the conditional void correlation of high-redshift mock quasars follows the 
hierarchical scaling but, in order to fully demonstrate the consistency of our model, we 
should explicitly check that this remains the case when we weight the sources according to 
Eq.(\ref{eq:weightJ}). We intend to test this assumption in a future work. 

\section{Conclusion}
\label{sec:conclusion}

We have developed a count-in-cell approach to the distribution of ionising intensity which 
includes source clustering. We have applied our method to quantify the impact of quasar 
clustering on the distribution of {\small HeII} ionising radiation at the end of helium 
reionization ($z\sim 3$). Our results can be summarised as follows:
\begin{itemize}
\item Our approach crucially relies on the assumption that the distribution of ionising sources 
follows the hierarchical ansatz. We have tested this hypothesis using catalogues of synthetic 
quasars at $z\sim 3$. We have found that the void scaling function of these mocks closely tracks 
the Negative Binomial scaling. Therefore, we have assumed that the real quasars follow the same 
pattern in all our predictions.
\item We have derived asymptotic expectations in the low- and high-intensity regime. We have 
shown that source clustering can noticeably increase the probability of finding ionising 
intensities $j\ll \langle j\rangle$, while it enhances the amplitude of the power-law tail
$\propto j^{-5/2}$ for $j\gtrsim \langle j\rangle$. We have implemented the numerical computation 
of the intensity distribution in two different ways to check the robustness of our numerical results.
\item Using the observationally determined quasar luminosity function and 2-point correlation, 
and ignoring any possible luminosity-dependence of quasar clustering, we have found that, for a 
(comoving) attenuation length in the range $25<r_0<55$ Mpc, quasar clustering becomes significant 
when the correlation length exceeds $\sim 15 - 20$ Mpc. Overall, the importance of source 
clustering increases with $\bar{N}_0\sim (r_0/l)^3$ (smaller Poisson fluctuations) and with 
$r_\xi/r_0$ (larger clustering strength).
\item We have shown that the distribution of ionising intensity depends on the surrounding 
environment. Variations of a few in the mean specific intensity $\langle j\rangle$ are expected 
for  large-scale ($R\sim 30$ Mpc), $\pm 2\sigma$ density fluctuations. However, a better 
characterisation of the connection between quasars and strong absorption systems is in order to
make more accurate predictions.
\end{itemize}
To conclude, quasar clustering is certainly not the dominant source of fluctuations in the
distribution of {\small HeII}-ionising intensity at $z\sim 3$. However, owing to the large 
uncertainties in the attenuation length $r_0$ and the clustering length $r_\xi$, it is difficult 
to draw any firm conclusion about the magnitude of this effect. If $r_\xi \lesssim 15$ Mpc and 
$r_0\lesssim 55$ Mpc, then quasar clustering is definitely a secondary source of variance (with 
a contribution less than $\sim$30\%), in agreement with the findings of 
\cite{dixon/furlanetto/mesinger:2014}. 
By contrast, if the attenuation length is on the high side of the allowed range, $r_0\gtrsim 55$ 
Mpc, and/or if the clustering of high-redshift quasars has a strong luminosity-dependence, with 
the rare bright quasars being highly clustered, then the variance of intensity fluctuations may 
be enhanced quite significantly. 

\section*{Acknowledgments}

We are grateful to Darren Croton for making his mock quasar catalogues available to us.
VD would like to thank the organisers of the Gravasco trimester at the Institute Henri Poincar\'e 
for hospitality when parts of this work were being completed; as well as St\'ephane Colombi, 
Ravi Sheth and Patrick Valageas for interesting discussions; and Sandrine Codis for correspondence.
MB would like to thank Alba Grassi for useful discussions.
VD and MB acknowledge support by Swiss National Science Foundation. 

\appendix

\section{Generating functional for weighted probabilities}
\label{app:CIC}

We begin with the probability to have a cell of volume $V$ empty of particles except at $N\geq 0$ 
distinct locations $\vx_1$, ... , $\vx_N$, 
\begin{align}
P\{\Phi_0(V)\} &= \exp\!\left[{\cal W}_0(V)\right] \\
P\{X_1\Phi_0(V)\} &= \bar{n}{\cal W}_1(\vx_1;V)\dd^3\!\vx_1\,P\{\Phi_0(V)\} \nonumber \\
P\{X_1 X_2\Phi_0(V)\} &= \bar{n}^2\Bigl[{\cal W}_1(\vx_1;V){\cal W}_1(\vx_2;V) 
\nonumber \\
& \qquad +{\cal W}_2(\vx_1,\vx_2;V)\Bigr] \dd^3\!\vx_1 \dd^3\!\vx_2\,P\{\Phi_0(V)\} \nonumber \\
& \qquad \qquad \dots \nonumber 
\end{align}
Here, ${\cal W}_N$ is the $N$-point conditional correlation function \citep[see Eq.(7) of][]{white:1979}. 
Substituting these relations into the series expansion Eq.(\ref{eq:pw1}), we obtain 
\begin{align}
P_\omega(V) &= \biggl\{\bar{n}  \Bigl( \xi_{k+1} \star \omega \Bigr) 
+ \frac{\bar{n}^2}{2!} \biggl[\Bigl( \xi_{k+1} \star \omega\Bigr)^2 
+ \Bigl( \xi_{k+2} \star \omega^2 \Bigr)\biggr] \nonumber \\
& \quad + \frac{\bar{n}^3}{3!} \biggl[ \Bigl( \xi_{k+1} \star\omega\Bigr)^3
+ 3 \Bigl(\xi_{k+1}\star\omega\Bigr)\Bigl(\xi_{k+2}\star\omega^2\Bigr) \nonumber \\
& \qquad  +\Bigl(\xi_{k+3}\star\omega^3\Bigr)\biggr] + \dots \biggr\}\,
e^{{\cal W}_0(V)} \;,
\end{align}
where $(\xi_{k+i}\star\omega^i)$ is a shorthand notation for the infinite sum
\begin{multline}
\sum_{k=0}^\infty \frac{(-\bar{n})^k}{k!} \int_V\!\!\dd^3\!\vx_1\dots\int_V\!\!\dd^3\!\vx_{k+i}\\
\times \xi_{k+i}(\vx_1,\dots,\vx_{k+i})\,\omega(\vx_1)\dots\omega(\vx_i)
\end{multline}
Note that this series involves irreducible correlation functions $\xi_N$ with $N\geq i$
solely.
Let us first consider a random distribution, for which $\xi_1(\vx)\equiv 1$ only is non-zero. 
Since $\xi_1(\vx)$ appears exclusively in $(\xi_{k+1}\star\omega)$, $P_\omega(V)$ simplifies to
\begin{align}
P_\omega(V) &= \biggl\{\bar{n}\int_V\!\!\dd^3\!\vx\,\omega(\vx) + 
\frac{\bar{n}^2}{2!} \left(\int_V\!\!\dd^3\!\vx\,\omega(\vx)\right)^2 
+ \dots \biggr\} e^{{\cal W}_0(V)} \nonumber \\
\label{eq:pomega_ran}
&= e^{-\bar{n}\int_V\!\dd^3\!\vx\,(1-\omega(\vx))}-e^{-\bar{n}V} \;.
\end{align}
When the distribution is not Poisson, the first non-trivial correlation is $\xi_2(\vx_1,\vx_2)$,
which contributes two terms at leading (linear) order,
\begin{multline}
-\bar{n}^2\int_V\!\!\dd^3\!\vx_1\int_V\!\!\dd^3\!\vx_2\,\xi_2(\vx_1,\vx_2)\omega(\vx_1) \\
+\frac{\bar{n}^2}{2!} \int_V\!\!\dd^3\!\vx_1\int_V\!\!\dd^3\!\vx_1\,\xi_2(\vx_1,\vx_2)
\omega(\vx_1)\omega(\vx_2) \;.
\end{multline}
The first arises from $\bar{n}(\xi_{k+1}\star\omega)$ whereas the second appears in
$(\bar{n}^2/2)(\xi_{k+2}\star\omega^2)$. This can also be written as
\begin{multline}
\frac{\bar{n}^2}{2!}\int_V\!\!\dd^3\!\vx_1\int_V\!\!\dd^3\!\vx_2\,
\xi_2(\vx_1,\vx_2) \\ 
\times \Bigl[\omega(\vx_1)\omega(\vx_2)-\omega(\vx_1)-\omega(\vx_2)\Bigr] \;.
\end{multline}
Furthermore, Taylor expanding the void probability $P_0=\exp[{\cal W}_0(V)]$ brings down an 
additional factor of
\begin{equation}
\frac{\bar{n}^2}{2!}\int_V\!\!\dd^3\!\vx_1\int_V\!\!\dd^3\!\vx_2\,\xi_2(\vx_1,\vx_2)\;,
\end{equation}
Hence, the terms linear in $\xi_2(\vx_1,\vx_2)$ are all quadratic in $\bar{n}$ and sum up to
\begin{equation}
\frac{\bar{n}^2}{2!}\int_V\!\!\dd^3\!\vx_1\int_V\!\!\dd^3\!\vx_2\,
\xi_2(\vx_1,\vx_2)\bigl(1-\omega(\vx_1)\bigr)\bigl(1-\omega(\vx_2)\bigr)\;.
\end{equation}
This suggests replacing Eq.(\ref{eq:pomega_ran}) by Eq.(\ref{eq:pomega}), where 
${\cal W}_\omega(V)$ is given by Eq.(\ref{eq:womega}). Subsequent checks at third order show that 
this must be the exact result. 

\section{Asymptotics}
\label{app:asymptotics}

\subsection{Low-intensity tail}
\label{app:low}

To estimate how $P(j)$ scales in the low-intensity limit, Eq.(\ref{eq:saddle}), we need 
to evaluate $F(j)$ and $F''(j)$. To this purpose, we follow \cite{meiksin/white:2003} and 
write the function $h(z)$ as
\begin{equation}
h(z)=\int_0^\infty\!\!\dd u\,e^{zu} \tau^3(u) \;,
\end{equation}
where $\tau(u)$ is solution of $u=\tau^{-2}e^{-\tau}$. On the negative real axis, only the
domain $0<u\ll 1$ contributes significantly to the integral in the limit $x\to -\infty$.
Hence, $\tau(u)\approx -\ln(u)$ and the function $h(x)$ is approximately
\begin{equation}
\label{eq:smallg}
h(x) \approx -\int_0^\infty\!\!\dd u\,e^{xu} \left(\ln u\right)^3 
\approx -\frac{\bigl(\ln(-x)\bigr)^3}{x}\;.
\end{equation}
The last equality is obtained upon retaining the dominant term to the integral solely. 
Therefore, we can approximate $G(x)$ as
\begin{align}
G(x) &\approx -\bar{N}_0 \int_{\alpha_{\rm min}}^{\alpha_{\rm max}}\!\!\dd\alpha\,\phi(\alpha)
\ln^3(-\alpha x) \\
&\sim -\bar{N}_0\ln^3(-\langle\alpha\rangle x) \nonumber \;,
\end{align}
so that its derivative reads $G'(x) \approx -3\bar{N}_0\ln^2(-\langle\alpha\rangle x)/x$. 
The condition $G'(x)=j$ leads to $x\approx -3\bar{N}_0/j$ for $j\ll 1$. Substituting this 
relation into the Legendre transform $F(j)=jx(j)-G\bigl[x(j)\bigr]$, we arrive at 
$F(j)\approx -\bar{N}_0 \bigl(\ln(j/j_c)\bigr)^3$.
We also find $F''(j)\approx 3\bar{N}_0 \bigl( j\ln(j/j_c)\bigr)^{-2}$. 
On inserting these expressions into Eq.(\ref{eq:saddle}), we obtain
\begin{equation}
P(j)\approx -\sqrt{\frac{3\bar{N}_0}{2\pi}}\, \bigl(j\ln(j/j_c)\bigr)^{-1} \,
e^{\bar{N}_0\ln^3(j/j_c)}\;,
\end{equation}
which leads to Eq.(\ref{eq:saddle1}) after multiplication by $j$.

In the presence of source clustering, the behaviour of ${\cal G}(x)=G(x)\chi(x)$ in the limit 
$x\to -\infty$ strongly depends on the average clustering strength $(\bar{N}_e\bar{\xi}_2)\!(x)$,
which is of positive sign on the whole negative real axis. For large cells, the integral over
one of the position vectors drops out and we are left with
\begin{align}
\left(\bar{N}_e\bar{\xi}_2\right)\!\!(x)  &\approx 
(-1)^\gamma\left(\frac{3\bar{N}_0\tau_\xi^\gamma}{3-\gamma}\right)x
\int_{\alpha_{\rm min}}^{\alpha_{\rm max}}\!\!\dd\alpha\,\alpha\phi(\alpha) \\ 
& \quad \times \int_0^{\infty}\!\!\dd u\, e^{\alpha x u} (\ln u)^{3-\gamma} \nonumber
\end{align}
in the limit $x\to -\infty$. We have assumed a power-law correlation with logarithmic slope 
$\gamma$. For $\gamma=2$ close to the observed value, the integral over the variable $u$ can be 
evaluated analytically~:
\begin{equation}
\int_0^\infty\!\!\dd u\, e^{\alpha xu}\left(\ln u\right)^{3-\gamma}
\approx \frac{\ln\bigl(-\alpha x\bigr)}{\alpha x} \;.
\end{equation}
Therefore, the average clustering strength scales according to
\begin{equation}
\left(\bar{N}_e\bar{\xi}_2\right)\!\!(x)\sim 
3 \bar{N}_0 \tau_\xi^2 \ln\left(-\langle\alpha\rangle x\right) \;,
\end{equation}
i.e. it diverges in the limit $x\to -\infty$. Since the void scaling function scales like 
$\chi\sim (\bar{N}_e\bar{\xi}_2)^{-1}$ for large average clustering strengths, we can approximate
${\cal G}(x)$ as
\begin{equation}
{\cal G}(x) \sim \frac{G(x)}{\left(\bar{N}_e\bar{\xi}_2\right)\!\!(x)}
\sim -\frac{1}{3\tau_\xi^2}\ln^2\bigl(-\langle\alpha\rangle x\bigr) \;. 
\end{equation}
Following the steps outlined above, we eventually find that the intensity distribution $P(j)$ 
scales as
\begin{equation}
P(j) \sim 
\exp\left[-\frac{1}{3\tau_\xi^2}\ln^2\!\left(\frac{3\tau_\xi^2 j}{2\langle\alpha\rangle}\right)\right]
\end{equation}
in the regime $j\ll j_c$. Note that this expression does not involve the mean number density $\bar{n}$ 
of the sources.

\subsection{High-intensity tail}
\label{app:high}

Consider the relation $F(j)+G(z)=jz$ which defines $F$ and $G$ as Legendre transforms. Let 
$f_n\equiv F^{(n)}(j_c)$ and $g_n\equiv G^{(n)}(z_c)$ denote derivatives of $F$ and $G$ at 
the critical point $(z_c,j_c)=(0,\left\langle j\right\rangle)$. 
Since $g_2$ is singular, the relation $g_2 f_2 = 1$ requires $f_2\equiv 0$. In addition, 
$f_1=z_c\equiv 0$ and $F(j_c)=j_c z_c - G(z_c)=0$ since $G(z_c)=0$. 
Therefore, a Taylor development of $F(j)$ and $F'(j)$ around $j_c$ yields
\begin{align}
F(j) &= \frac{1}{6}f_3\left(j-j_c\right)^3 +\dots \\
F'(j) &= \frac{1}{2} f_3\left(j-j_c\right)^2 + \dots
\end{align} 
The bottom panel of Fig.\ref{fig:legendre} demonstrates that $f_3$ must be a negative real 
number. On writing $F'(j)=z$, the second relation can be inverted to obtain $j(z)$, i.e.
\begin{equation}
j-j_c = -\sqrt{\frac{2}{f_3}}\, z^{1/2}\;.
\end{equation}
The minus sign ensures that $j<j_c$ when $z<0$. Hence, in the vicinity of $z_c=0$, the function
$G(z)$ reads
\begin{align}
\label{eq:Gzseries}
G(z) &\equiv z j(z) - F\bigl[j(z)\bigr] \\
&= z j_c - \frac{2}{3}\sqrt{\frac{2}{f_3}} \,z^{3/2} + \dots \nonumber
\end{align} 
which leads to Eq.(\ref{eq:argzc}). We then must expand $G(z)$ in the limit $z\to -\infty$.

The integral along the contour $C$ is the sum of two contributions $I_1+I_2$, where $I_1$ is 
the integral along the semi-circular contour of radius $\epsilon$ centred at $z=0$ and
\begin{equation}
I_2 = \frac{1}{2\pi i}
\left(\int_{-i\epsilon+\infty}^{-i\epsilon}\!\!\dd z+\int_{i\epsilon}^{i\epsilon+\infty}\!\!\dd z\right)
\,e^{-jz + G(z)} \;.
\end{equation}
Since $G(z^\star)=G(z)^\star$, the first integral is equal to minus the complex conjugate of the
second. Furthermore, $I_1\to 0$ in the limit $\epsilon \to 0$. Therefore,
\begin{equation}
P(j)= \frac{1}{\pi}{\rm Im}\left\{\int_{i\epsilon}^{i\epsilon+\infty}\!\!\dd z
\,\,e^{-jz + G(z)}\right\} \;.
\end{equation}
Substituting the series Eq.(\ref{eq:argzc}) into the argument of the exponential and taking the 
limit $\epsilon\to 0$, we have
\begin{align}
P(j) &\approx \frac{1}{\pi} {\rm Im}\left\{\int_0^\infty\!\!\dd x\, e^{-(j-j_c)x}
\left[1 -\frac{2}{3}\sqrt{\frac{2}{f_3}}x^{3/2}+\dots \right]\right\} \nonumber  \\
& = -\frac{1}{\pi}{\rm Im}\left\{\int_0^\infty \!\!\dd x\,e^{-(j-j_c)x}\frac{2}{3}
\sqrt{\frac{2}{f_3}} x^{3/2} \right\} \;.
\end{align}
Performing the integral over $x$ (which is proportional to $\Gamma(5/2)$), we arrive at 
Eq.(\ref{eq:pjasym1}).

The last step of the calculation is the evaluation of $f_3$. Eqs.~(\ref{eq:Gz}) and (\ref{eq:Gzseries}) 
show that the function $h(z)$ must admit the series expansion  
\begin{equation}
h(z)= 3 + c_{1/2} z^{1/2} + c_1 z + \dots
\end{equation}
in the vicinity of the critical point $z=z_c=0$. A numerical evaluation of $(h(x) - 3)x^{-1/2}$ 
in the limit $x\to 0$ yields $c_{1/2}=2\sqrt{\pi}i$. 
The factor of $i$ ensures that $h(x)$ is real on the negative real axis. Therefore, 
\begin{align}
G(z) &= z \bar{N}_0 
\int_{\alpha_{\rm min}}^{\alpha_{\rm max}}\!\!\dd\alpha\,\alpha\,\phi(\alpha) 
\left(3+2\sqrt{\pi}i\alpha^{1/2}z^{1/2}+\dots\right) \nonumber \\
&\approx z j_c + 2 \sqrt{\pi} i \bar{N}_0 \bigl\langle\alpha^{3/2}\bigr\rangle z^{3/2} \;,
\end{align}
from which we easily read off the value of $f_3$~:
\begin{equation}
\label{eq:f3}
f_3 = -\frac{2}{9\pi\bar{N}_0^2\bigl\langle\alpha^{3/2}\bigr\rangle^2} < 0 \;.
\end{equation}
Substituting this expression into Eq.(\ref{eq:pjasym1}) yields the desired expression
Eq.(\ref{eq:pjasym2}).

In the presence of source clustering, we must consider the behaviour of the average clustering
strength in the neighbourhood of $z=0$. For a power-law correlation function, we have
\begin{align}
\left(\bar{N}_e\bar{\xi}_2\right)\!\!(z) &=
-\left(\frac{3\bar{N}_0\tau_\xi^\gamma}{3-\gamma}\right)z
\int_{\alpha_{\rm min}}^{\alpha_{\rm max}}\!\!\dd\alpha\,\alpha\,\phi(\alpha) \nonumber \\
&\qquad \times 
\int_0^\infty\!\!\dd\tau\,\tau^{-\gamma}e^{z\alpha \tau^{-2}e^{-\tau}}e^{-\tau}\left(2+\tau\right)
\end{align}
in the limit of large cells. Unlike $h(z)$, the integral over $\tau$ diverges in the limit 
$z\to 0$. This can be seen upon substituting the variable $w=\tau^{-2}e^{-\tau}$ as in 
\cite{meiksin/white:2003}. The $\tau$-integral then becomes
\begin{align}
\int_0^\infty\!\!\dd w\,\tau(w)^{3-\gamma}e^{z\alpha w}
&\approx \int_0^\infty\!\!\dd w\,w^{-(3-\gamma)/2}e^{-(-z\alpha)w} \nonumber \\
&= (-z\alpha)^{(1-\gamma)/2}\,\Gamma\left(\frac{\gamma-1}{2}\right) \;.
\label{eq:zdivergence}
\end{align}
The first equality follows from the assumption $\tau=w^{-1/2}$, which is a very good approximation
at the small optical depths responsible for the divergence of the integral. For $\gamma\approx 2$, 
the integral diverges as $z^{-1/2}$. 
Therefore, applying the above Legendre transform to $(\bar{N}_e\bar{\xi}_2)(z)$ rather than $G(z)$
suggests that the average clustering strength admits the series expansion
\begin{equation}
\left(\bar{N}_e\bar{\xi}_2\right)\!\!(z) = c_{1/2} z^{1/2} + c_1 z + c_{3/2} z^{3/2} + \dots
\end{equation}
Eq.(\ref{eq:zdivergence}) provides an estimate for the coefficient of the leading term,
\begin{equation}
c_{1/2} \approx -3\sqrt{\pi}i\tau_\xi^2\bar{N}_0\bigl\langle\alpha^{1/2}\bigr\rangle \;.
\end{equation}
Here, the minus sign ensures that $\bar{N}_e\bar{\xi}_2>0$ when $z$ is on the negative real axis. 
Using the fact that the void scaling function always behaves like
$\chi(\bar{N}_e\bar{\xi}_2)\approx 1-(1/2)\bar{N}_e\bar{\xi}_2$ for small clustering strength, we
arrive at
\begin{align}
\label{eq:highasym}
{\cal G}(z) &= G(z) \chi(z) \\
&\approx z j_c + 2\sqrt{\pi}i\bar{N}_0\bigl\langle\alpha^{3/2}\bigr\rangle
\left(1+\frac{9}{4}\tau_\xi^2\bar{N}_0\frac{\langle\sqrt{\alpha}\rangle\langle\alpha\rangle}
{\langle\alpha^{3/2}\rangle}\right) z^{3/2} \nonumber \;.
\end{align}
This implies that source clustering enhances the amplitude of the high-intensity tail $\propto j^{-5/2}$
by the factor given in the parenthesis. This enhancement is indeed observed in Fig.\ref{fig:pj1},
albeit with a $\sim 6$ times smaller amplitude.

\bibliographystyle{mn2e}
\bibliography{references}

\begin{thebibliography}{74}
\expandafter\ifx\csname natexlab\endcsname\relax\def\natexlab#1{#1}\fi

\bibitem[{{Balian} \& {Schaeffer}(1989)}]{balian/schaeffer:1989}
{Balian} R., {Schaeffer} R., 1989, \aap, 220, 1

\bibitem[{{Bernardeau}(1994)}]{bernardeau:1994}
{Bernardeau} F., 1994, \aap, 291, 697

\bibitem[{{Bernardeau} \& {Kofman}(1995)}]{bernardeau/kofman:1995}
{Bernardeau} F., {Kofman} L., 1995, \apj, 443, 479

\bibitem[{{Bernardeau}, {Pichon} \& {Codis}(2013){Bernardeau}, {Pichon}, \&
  {Codis}}]{bernardeau/codis/pichon:2013}
{Bernardeau} F., {Pichon} C., {Codis} S., 2013, ArXiv e-prints

\bibitem[{{Bernardeau} \& {Schaeffer}(1999)}]{bernardeau/schaeffer:1999}
{Bernardeau} F., {Schaeffer} R., 1999, \aap, 349, 697

\bibitem[{{Berntsen}, {Espelid} \& {Genz}(1991){Berntsen}, {Espelid}, \&
  {Genz}}]{berntsen/espelid/genz:1991}
{Berntsen} J., {Espelid} T., {Genz} A., 1991, ACM Transactions on Mathematical
  Software, 17, 452

\bibitem[{{Biagetti} {et~al}\mbox{.}(2014){Biagetti}, {Chan}, {Desjacques}, \&
  {Paranjape}}]{biagetti/chan/etal:2014}
{Biagetti} M., {Chan} K.~C., {Desjacques} V., {Paranjape} A., 2014, \mnras,
  441, 1457

\bibitem[{{Bolton} {et~al}\mbox{.}(2006){Bolton}, {Haehnelt}, {Viel}, \&
  {Carswell}}]{bolton/haehnelt/etal:2006}
{Bolton} J.~S., {Haehnelt} M.~G., {Viel} M., {Carswell} R.~F., 2006, \mnras,
  366, 1378

\bibitem[{{Bouchet} {et~al}\mbox{.}(1993){Bouchet}, {Strauss}, {Davis},
  {Fisher}, {Yahil}, \& {Huchra}}]{bouchet/strauss/etal:1993}
{Bouchet} F.~R., {Strauss} M.~A., {Davis} M., {Fisher} K.~B., {Yahil} A.,
  {Huchra} J.~P., 1993, \apj, 417, 36

\bibitem[{{Boyle} {et~al}\mbox{.}(1993){Boyle}, {Griffiths}, {Shanks},
  {Stewart}, \& {Georgantopoulos}}]{boyle/griffiths/etal:1993}
{Boyle} B.~J., {Griffiths} R.~E., {Shanks} T., {Stewart} G.~C.,
  {Georgantopoulos} I., 1993, \mnras, 260, 49

\bibitem[{{Boyle}, {Shanks} \& {Peterson}(1988){Boyle}, {Shanks}, \&
  {Peterson}}]{boyle/shanks/peterson:1988}
{Boyle} B.~J., {Shanks} T., {Peterson} B.~A., 1988, \mnras, 235, 935

\bibitem[{{Carruthers} \& {Shih}(1983)}]{carruthers/shih:1983}
{Carruthers} P., {Shih} C.~C., 1983, Physics Letters B, 127, 242

\bibitem[{{Colombi} {et~al}\mbox{.}(1997){Colombi}, {Bernardeau}, {Bouchet}, \&
  {Hernquist}}]{colombi/bernardeau/etal:1997}
{Colombi} S., {Bernardeau} F., {Bouchet} F.~R., {Hernquist} L., 1997, \mnras,
  287, 241

\bibitem[{{Colombi}, {Bouchet} \& {Schaeffer}(1995){Colombi}, {Bouchet}, \&
  {Schaeffer}}]{colombi/bouchet/schaeffer:1995}
{Colombi} S., {Bouchet} F.~R., {Schaeffer} R., 1995, \apjs, 96, 401

\bibitem[{{Conroy} \& {White}(2013)}]{conroy/white:2013}
{Conroy} C., {White} M., 2013, \apj, 762, 70

\bibitem[{{Croom} {et~al}\mbox{.}(2004){Croom}, {Smith}, {Boyle}, {Shanks},
  {Miller}, {Outram}, \& {Loaring}}]{croom/smith/etal:2004}
{Croom} S.~M., {Smith} R.~J., {Boyle} B.~J., {Shanks} T., {Miller} L., {Outram}
  P.~J., {Loaring} N.~S., 2004, \mnras, 349, 1397

\bibitem[{{Croton}(2009)}]{croton:2009}
{Croton} D.~J., 2009, \mnras, 394, 1109

\bibitem[{{Croton} {et~al}\mbox{.}(2004){Croton}, {Gazta{\~n}aga}, {Baugh},
  {Norberg}, {Colless}, {Baldry}, {Bland-Hawthorn}, {Bridges}, {Cannon},
  {Cole}, {Collins}, {Couch}, {Dalton}, {De Propris}, {Driver}, {Efstathiou},
  {Ellis}, {Frenk}, {Glazebrook}, {Jackson}, {Lahav}, {Lewis}, {Lumsden},
  {Maddox}, {Madgwick}, {Peacock}, {Peterson}, {Sutherland}, \&
  {Taylor}}]{croton/gaztanaga/eta:2004}
{Croton} D.~J. {et~al.}, 2004, \mnras, 352, 1232

\bibitem[{{Davies} \& {Furlanetto}(2014)}]{davies/furlanetto:2014}
{Davies} F.~B., {Furlanetto} S.~R., 2014, \mnras, 437, 1141

\bibitem[{{Desjacques}, {Nusser} \& {Sheth}(2007){Desjacques}, {Nusser}, \&
  {Sheth}}]{desjacques/nusser/sheth:2007}
{Desjacques} V., {Nusser} A., {Sheth} R.~K., 2007, \mnras, 374, 206

\bibitem[{{Dixon} \& {Furlanetto}(2009)}]{dixon/furlanetto:2009}
{Dixon} K.~L., {Furlanetto} S.~R., 2009, \apj, 706, 970

\bibitem[{{Dixon}, {Furlanetto} \& {Mesinger}(2014){Dixon}, {Furlanetto}, \&
  {Mesinger}}]{dixon/furlanetto/mesinger:2014}
{Dixon} K.~L., {Furlanetto} S.~R., {Mesinger} A., 2014, \mnras, 440, 987

\bibitem[{{Djorgovski} {et~al}\mbox{.}(2003){Djorgovski}, {Stern}, {Mahabal},
  \& {Brunner}}]{djorgovski/stern/etal:2003}
{Djorgovski} S.~G., {Stern} D., {Mahabal} A.~A., {Brunner} R., 2003, \apj, 596,
  67

\bibitem[{{Fall} {et~al}\mbox{.}(1976){Fall}, {Geller}, {Jones}, \&
  {White}}]{fall/geller/etal:1976}
{Fall} S.~M., {Geller} M.~J., {Jones} B.~J.~T., {White} S.~D.~M., 1976, \apjl,
  205, L121

\bibitem[{{Fardal} \& {Shull}(1993)}]{fardall/shull:1993}
{Fardal} M.~A., {Shull} J.~M., 1993, \apj, 415, 524

\bibitem[{{Faucher-Gigu{\`e}re} {et~al}\mbox{.}(2009){Faucher-Gigu{\`e}re},
  {Lidz}, {Zaldarriaga}, \& {Hernquist}}]{fauchergiguere/lidz/etal:2009}
{Faucher-Gigu{\`e}re} C.-A., {Lidz} A., {Zaldarriaga} M., {Hernquist} L., 2009,
  \apj, 703, 1416

\bibitem[{{Font-Ribera} {et~al}\mbox{.}(2012){Font-Ribera},
  {Miralda-Escud{\'e}}, {Arnau}, {Carithers}, {Lee}, {Noterdaeme}, {P{\^a}ris},
  {Petitjean}, {Rich}, {Rollinde}, {Ross}, {Schneider}, {White}, \&
  {York}}]{font-ribera/miralda-escude/etal:2012}
{Font-Ribera} A. {et~al.}, 2012, \jcap, 11, 59

\bibitem[{{Francke} {et~al}\mbox{.}(2008){Francke}, {Gawiser}, {Lira},
  {Treister}, {Virani}, {Cardamone}, {Urry}, {van Dokkum}, \&
  {Quadri}}]{francke/gawiser/etal:2008}
{Francke} H. {et~al.}, 2008, \apjl, 673, L13

\bibitem[{{Fry}(1985)}]{fry:1985}
{Fry} J.~N., 1985, \apj, 289, 10

\bibitem[{{Fry}(1986)}]{fry:1986a}
{Fry} J.~N., 1986, \apj, 306, 358

\bibitem[{{Fry} \& {Colombi}(2013)}]{fry/colombi:2013}
{Fry} J.~N., {Colombi} S., 2013, \mnras, 433, 581

\bibitem[{{Fry} {et~al}\mbox{.}(2011){Fry}, {Colombi}, {Fosalba}, {Balaraman},
  {Szapudi}, \& {Teyssier}}]{fry/colombi/etal:2011}
{Fry} J.~N., {Colombi} S., {Fosalba} P., {Balaraman} A., {Szapudi} I.,
  {Teyssier} R., 2011, \mnras, 415, 153

\bibitem[{{Furlanetto}(2009)}]{furlanetto:2008}
{Furlanetto} S.~R., 2009, \apj, 703, 702

\bibitem[{{Furlanetto} \& {Oh}(2008)}]{furlanetto/oh:2008}
{Furlanetto} S.~R., {Oh} S.~P., 2008, \apj, 681, 1

\bibitem[{{Gaztanaga}(1994)}]{gaztanaga:1994}
{Gaztanaga} E., 1994, \mnras, 268, 913

\bibitem[{{Gleser} {et~al}\mbox{.}(2005){Gleser}, {Nusser}, {Benson}, {Ohno},
  \& {Sugiyama}}]{gleser/nusser/etal:2005}
{Gleser} L., {Nusser} A., {Benson} A.~J., {Ohno} H., {Sugiyama} N., 2005,
  \mnras, 361, 1399

\bibitem[{{Gunn} \& {Gott}(1972)}]{gunn/gott:1972}
{Gunn} J.~E., {Gott}, III J.~R., 1972, \apj, 176, 1

\bibitem[{{Haiman} \& {Hui}(2001)}]{haiman/hui:2001}
{Haiman} Z., {Hui} L., 2001, \apj, 547, 27

\bibitem[{{Hamilton}(1988)}]{hamilton:1988}
{Hamilton} A.~J.~S., 1988, \apj, 332, 67

\bibitem[{{Hennawi} {et~al}\mbox{.}(2006){Hennawi}, {Strauss}, {Oguri},
  {Inada}, {Richards}, {Pindor}, {Schneider}, {Becker}, {Gregg}, {Hall},
  {Johnston}, {Fan}, {Burles}, {Schlegel}, {Gunn}, {Lupton}, {Bahcall},
  {Brunner}, \& {Brinkmann}}]{hennawi/strauss/etal:2006}
{Hennawi} J.~F. {et~al.}, 2006, \aj, 131, 1

\bibitem[{{Hopkins}, {Richards} \& {Hernquist}(2007){Hopkins}, {Richards}, \&
  {Hernquist}}]{hopkins/etal:2007}
{Hopkins} P.~F., {Richards} G.~T., {Hernquist} L., 2007, \apj, 654, 731

\bibitem[{{Kaiser}(1984)}]{kaiser:1984}
{Kaiser} N., 1984, \apjl, 284, L9

\bibitem[{{Kundi{\'c}}(1997)}]{kundic:1997}
{Kundi{\'c}} T., 1997, \apj, 482, 631

\bibitem[{{Lahav} \& {Saslaw}(1992)}]{lahav/saslaw:1992}
{Lahav} O., {Saslaw} W.~C., 1992, \apj, 396, 430

\bibitem[{{Lepage}(1978)}]{lepage:1978}
{Lepage} G.~P., 1978, Journal of Computational Physics, 27, 192

\bibitem[{{Madau}, {Haardt} \& {Rees}(1999){Madau}, {Haardt}, \&
  {Rees}}]{madau/haardt/rees:1999}
{Madau} P., {Haardt} F., {Rees} M.~J., 1999, \apj, 514, 648

\bibitem[{{Martini} \& {Weinberg}(2001)}]{martini/weinberg:2001}
{Martini} P., {Weinberg} D.~H., 2001, \apj, 547, 12

\bibitem[{{Maselli} \& {Ferrara}(2005)}]{maselli/ferrara:2005}
{Maselli} A., {Ferrara} A., 2005, \mnras, 364, 1429

\bibitem[{{Meiksin} \& {Tittley}(2012)}]{meiksin/tittley:2012}
{Meiksin} A., {Tittley} E.~R., 2012, \mnras, 423, 7

\bibitem[{{Meiksin} \& {White}(2003)}]{meiksin/white:2003}
{Meiksin} A., {White} M., 2003, \mnras, 342, 1205

\bibitem[{{Mo} \& {White}(1996)}]{mo/white:1996}
{Mo} H.~J., {White} S.~D.~M., 1996, \mnras, 282, 347

\bibitem[{{Myers} {et~al}\mbox{.}(2008){Myers}, {Richards}, {Brunner},
  {Schneider}, {Strand}, {Hall}, {Blomquist}, \&
  {York}}]{myers/richards/etal:2008}
{Myers} A.~D., {Richards} G.~T., {Brunner} R.~J., {Schneider} D.~P., {Strand}
  N.~E., {Hall} P.~B., {Blomquist} J.~A., {York} D.~G., 2008, \apj, 678, 635

\bibitem[{{Padmanabhan} {et~al}\mbox{.}(2009){Padmanabhan}, {White}, {Norberg},
  \& {Porciani}}]{padmanabhan/white/etal:2009}
{Padmanabhan} N., {White} M., {Norberg} P., {Porciani} C., 2009, \mnras, 397,
  1862

\bibitem[{{Paschos} {et~al}\mbox{.}(2007){Paschos}, {Norman}, {Bordner}, \&
  {Harkness}}]{paschos/norman/etal:2007}
{Paschos} P., {Norman} M.~L., {Bordner} J.~O., {Harkness} R., 2007, ArXiv
  e-prints

\bibitem[{{Peebles}(1980)}]{peebles:1980}
{Peebles} P.~J.~E., 1980, {The large-scale structure of the universe},
  {Peebles, P.~J.~E.}, ed.

\bibitem[{{Pei}(1995)}]{pei:1995}
{Pei} Y.~C., 1995, \apj, 438, 623

\bibitem[{{Porciani}, {Magliocchetti} \& {Norberg}(2004){Porciani},
  {Magliocchetti}, \& {Norberg}}]{porciani/magliocchetti/norberg:2004}
{Porciani} C., {Magliocchetti} M., {Norberg} P., 2004, \mnras, 355, 1010

\bibitem[{{Prochaska} {et~al}\mbox{.}(2014){Prochaska}, {Madau}, {O'Meara}, \&
  {Fumagalli}}]{prochaska/madau/etal:2014}
{Prochaska} J.~X., {Madau} P., {O'Meara} J.~M., {Fumagalli} M., 2014, \mnras,
  438, 476

\bibitem[{{Ross}, {Brunner} \& {Myers}(2006){Ross}, {Brunner}, \&
  {Myers}}]{ross/brunner/myers:2006}
{Ross} A.~J., {Brunner} R.~J., {Myers} A.~D., 2006, \apj, 649, 48

\bibitem[{{Schneider} {et~al}\mbox{.}(2000){Schneider}, {Fan}, {Strauss},
  {Gunn}, {Richards}, {Knapp}, {Lupton}, {Saxe}, {Anderson}, {Bahcall},
  {Brinkmann}, {Brunner}, {Csabai}, {Fukugita}, {Hennessy}, {Hindsley},
  {Ivezi{\'c}}, {Nichol}, {Pier}, \& {York}}]{schneider/fan/etal:2000}
{Schneider} D.~P. {et~al.}, 2000, \aj, 120, 2183

\bibitem[{{Shen} {et~al}\mbox{.}(2007){Shen}, {Strauss}, {Oguri}, {Hennawi},
  {Fan}, {Richards}, {Hall}, {Gunn}, {Schneider}, {Szalay}, {Thakar}, {Vanden
  Berk}, {Anderson}, {Bahcall}, {Connolly}, \&
  {Knapp}}]{shen/strauss/etal:2007}
{Shen} Y. {et~al.}, 2007, \aj, 133, 2222

\bibitem[{{Sheth}(1996)}]{sheth:1996}
{Sheth} R.~K., 1996, \mnras, 278, 101

\bibitem[{{Sheth}(1998)}]{sheth:1998}
{Sheth} R.~K., 1998, \mnras, 300, 1057

\bibitem[{{Sokasian}, {Abel} \& {Hernquist}(2002){Sokasian}, {Abel}, \&
  {Hernquist}}]{sokasian/abel/hernquist:2002}
{Sokasian} A., {Abel} T., {Hernquist} L., 2002, \mnras, 332, 601

\bibitem[{{Springel} {et~al}\mbox{.}(2005){Springel}, {White}, {Jenkins},
  {Frenk}, {Yoshida}, {Gao}, {Navarro}, {Thacker}, {Croton}, {Helly},
  {Peacock}, {Cole}, {Thomas}, {Couchman}, {Evrard}, {Colberg}, \&
  {Pearce}}]{springel/white/etal:2005}
{Springel} V. {et~al.}, 2005, \nat, 435, 629

\bibitem[{{Stephens} {et~al}\mbox{.}(1997){Stephens}, {Schneider}, {Schmidt},
  {Gunn}, \& {Weinberg}}]{stephens/schneider/etal:1997}
{Stephens} A.~W., {Schneider} D.~P., {Schmidt} M., {Gunn} J.~E., {Weinberg}
  D.~H., 1997, \aj, 114, 41

\bibitem[{{Szapudi} \& {Colombi}(1996)}]{szapudi/colombi:1996}
{Szapudi} I., {Colombi} S., 1996, \apj, 470, 131

\bibitem[{{Szapudi} \& {Szalay}(1993)}]{szapudi/szalay:1993}
{Szapudi} I., {Szalay} A.~S., 1993, \apj, 408, 43

\bibitem[{{Telfer} {et~al}\mbox{.}(2002){Telfer}, {Zheng}, {Kriss}, \&
  {Davidsen}}]{telfer/zheng/etal:2002}
{Telfer} R.~C., {Zheng} W., {Kriss} G.~A., {Davidsen} A.~F., 2002, \apj, 565,
  773

\bibitem[{{Tittley} \& {Meiksin}(2007)}]{tittley/meiksin:2007}
{Tittley} E.~R., {Meiksin} A., 2007, \mnras, 380, 1369

\bibitem[{{Valageas}(2002)}]{valageas:2002}
{Valageas} P., 2002, \aap, 382, 412

\bibitem[{{Valageas} \& {Munshi}(2004)}]{valageas/munshi:2004}
{Valageas} P., {Munshi} D., 2004, \mnras, 354, 1146

\bibitem[{{White}(1979)}]{white:1979}
{White} S.~D.~M., 1979, \mnras, 186, 145

\bibitem[{{Zuo}(1992)}]{zuo:1992}
{Zuo} L., 1992, \mnras, 258, 36

\end{thebibliography}

\label{lastpage}

\end{document}